\documentstyle[12pt,bezier,worldsci]{article}
\makeatletter
\@addtoreset{equation}{section}
\makeatother
\def\prepno{DPNU-96-10 \\ hep-ph/9603293}%
\def\prepdate{March, 1996}

\newcommand{\VEV}[1]{\langle #1 \rangle}
\newcommand{\bra}[1]{\langle #1 |}
\newcommand{\ket}[1]{| #1 \rangle}

\newcommand{\fsl}[1]{\not \kern-2pt #1} 
\newcommand{\DISP}{\displaystyle}

\newcommand{\integ}[2]{\int_{#1}^{#2}\!\!}

\newcommand{\GeV}{\hbox{GeV}}
\newcommand{\MeV}{\hbox{MeV}}

\newcommand{\xbar}[1]{#1 \hspace{-5.5pt}/}

\newcommand{\SxSB}{\mbox{$S\chi SB$}}

\newcommand{\dfrac}[2]{\frac{\displaystyle{#1}}{\displaystyle{#2}}}
\def\fsl#1{\setbox0=\hbox{$#1$}           
   \dimen0=\wd0                                 
   \setbox1=\hbox{/} \dimen1=\wd1               
   \ifdim\dimen0>\dimen1                        
      \rlap{\hbox to \dimen0{\hfil/\hfil}}      
      #1                                        
   \else                                        
      \rlap{\hbox to \dimen1{\hfil$#1$\hfil}}   
      /                                         
   \fi}                                         %

\newcommand{\critline}{
\begin{picture}(150,150)(-15,-15)
   \thinlines
  \put(-10,0){\line(1,0){150}}
  \put(0,-10){\line(0,1){150}}
  \thicklines
  \put(100,-10){\line(0,1){35}}
  \put(107,-15){$\lambda_c$}
  \put(-2,100){\line(1,0){5}}
  \put(-15,107){1}
  \put(17,97){$g=g^{*}(\lambda)$}
  \bezier{300}(0,100)(99.7,50)(99.7,25)
  \put(25,25){$Sym.$}
  \put(75,75){\SxSB}
\end{picture}}

\title{DYNAMICAL SYMMETRY BREAKING WITH LARGE ANOMALOUS 
DIMENSION
\thanks{To appear in Proc. 14th Symposium on Theoretical Physics      
   ``Dynamical Symmetry Breaking and Effective Field Theory'',         
                     Cheju, Korea,                                     
                    July 21-26, 1995                                   
}\
\author{
  KOICHI YAMAWAKI\\
  Department of Physics, Nagoya University \\
  Nagoya 464-01, Japan
}
}
\date{}
\begin{document}

\maketitle

\begin{abstract}
  \small
  \setlength{\baselineskip}{12pt}
  We give an introduction to 
  the dynamical symmetry breaking 
  with large anomalous dimension. This is the basis of tightly
  bound composite Higgs models such as walking technicolor, strong ETC 
  technicolor and top quark condensate, which are all characterized by
   the large anomalous dimension, $\gamma_m \simeq 1$ (walking technicolor),
  $1<\gamma_m<2$ (strong ETC technicolor) and $\gamma_m \simeq 2$ (top quark 
  condensate) due to nontrivial short distance dynamics
  of the gauged Nambu-Jona-Lasinio (NJL)
  models (gauge theories plus four-fermion interactions). 

  Particular emphasis will be placed on the top quark condensate in which
  the critical phenomenon in the gauged NJL model 
  yields a simple reason why the top 
  quark can have an extremely large mass compared with other quarks and 
  leptons. Topics will also cover a recent observation that  
  the four-fermion theory {\it in the presence
  of gauge interactions} (gauged NJL model) 
  can become renormalizable and nontrivial  in sharp 
  contrast to the pure NJL model without gauge interactions.
  The requirement of this renormalizability/nontriviality
  of the gauged NJL model can be applied to
  the top quark condensate when the standard gauge
  groups are unified into a GUT with ``walking'' coupling, which  then
  naturally leads to the top quark and the Higgs masses both around 180 GeV. 

\end{abstract}

\tableofcontents

\section{Introduction}

As it stands now, the standard model (SM) is a very successful framework for 
describing elementary particles in the low energy region, say, less than 
100 GeV. However one of the most mysterious part of the theory, 
{\it the origin of mass}, has long been left unexplained. Actually, 
mass of {\it all} particles in the SM is attributed 
to a {\it single} order parameter, the vacuum expectation value (VEV) of 
the Higgs doublet. Thus the problem of the origin of mass is simply 
reduced to understanding the dynamics of the Higgs sector.

Here we note that 
the situation very much resembles the 
Ginzburg-Landau (GL)'s {\it macroscopic} theory for 
the superconductivity, the mysterious parts of which were 
eventually explained by the {\it microscopic} theory of 
Bardeen-Cooper-Schrieffer (BCS): 
The GL's phenomenological order parameter was replaced by 
the Cooper pair condensate due to the short range attractive forces.

A similar thing has also happened to the hadron physics where the $\sigma$ model 
description by Gell-Mann and Levy (GML) works very well as far as the low 
energy (macroscopic) phenomena are concerned, while the deeper understanding 
of it was first given by Nambu and Jona-Lasinio (NJL)\cite{kn:NJL61} 
based on the analogy with the BCS dynamics.
Nowadays people believe that essentially the same phenomena as described by 
the NJL paper takes place in the microscopic theory for hadrons, QCD. In QCD 
the VEV of $\sigma$, the GML's order parameter $\VEV\sigma=f_\pi=93\MeV$, has 
been replaced by the quark-antiquark pair condensate 
$\VEV{\bar{q} q}=O(f_\pi^3)$,
an analogue of the Cooper pair condensate, formed by the attractive color 
forces. 
The Nambu-Goldstone (NG) boson, the pion, is now a composite state of 
 quark and antiquark.
This is actually the prototype of the {\it dynamical symmetry breaking} 
(DSB) due to {\it composite order parameters} like fermion pair condensates.

In fact Higgs sector in the SM is precisely the same as the 
$\sigma$ model except that $\VEV\sigma=f_\pi=93\MeV$ is now replaced by the 
Higgs VEV$=F_\pi=250\GeV$, roughly a 2600 times scale-up. One is thus 
naturally led to speculate 
that there might exist a microscopic theory for the Higgs sector, with the 
Higgs VEV being replaced by the fermion-antifermion pair condensate due to 
yet another strong interaction called technicolor (TC)\cite{kn:Wein76}. 

Unfortunately, the original version of TC was too naive to survive 
the FCNC (flavor-changing neutral current) syndrome\cite{kn:FS81}.
It was not the end of the story, however. QCD-like theories 
(simple scale-up's of QCD) turned 
out {\it not} to be the {\it unique} candidate
for the underlying  composite dynamics of the Higgs sector. 

Actually, there have been
proposed a variety of composite Higgs models which, though equally behaving 
as the $\sigma$ model in the low energy region, still have different
high energy behaviors than QCD; walking technicolor\cite{kn:Hold85,kn:YBM86}, 
strong ETC technicolor\cite{kn:MY89,kn:ATEW89} and the top quark condensate
(top mode standard model)\cite{kn:MTY89a,kn:MTY89b,kn:Namb89,kn:Marc89,kn:BHL90}, etc.. 
Interactions in these models persist at  high energy or short distance
and hence produce tightly bound composite Higgs.

These tightly bound composite Higgs models were actually proposed based
on the explicit dynamics, 
gauged NJL model (gauge 
theory plus four-fermion interaction) within the framework of
ladder Schwinger-Dyson (SD) equation. 
The gauged NJL model\cite{kn:BLL86} was shown to have a phase structure 
divided by a critical coupling (critical line)\cite{kn:KMY89}
 similarly to the NJL model, 
 and have a {\it large anomalous 
 dimension} \cite{kn:MY89,kn:KSY91,kn:KTY93} due to {\it strong attractive 
 forces  at relatively short distance or high energy}. 
Such a system may actually be regarded as a 
theory with ultraviolet fixed point(s) in contrast to the 
asymptotic freedom.
A remarkable feature of this dynamics is that the four-fermion interation in 
four dimensions
may become {\it renormalizable} thanks to the large anomalous dimension 
($\gamma_m >1$) and the 
presence of gauge interaction ($\gamma_m <2$) in sharp contrast
to the pure NJL model with ``$\gamma_m =2$''.%
\cite{kn:KSY91,kn:KTY93,kn:Yama92,kn:Kras93,kn:KSTY94,kn:HKKN94}

In this lecture we would like
to give a general description of DSB with large anomalous dimension
and tightly bound composite Higgs models based on that dynamics.
Main parts of this subject have already been covered by many 
reviews\cite{kn:Yama92,kn:Yama89}
and a textbook\cite{kn:Mira94}.
Here we shall  put
a stress rather on the renormalizability of the gauged 
NJL model and its possible
application\cite{kn:ITY96} to modifications of the 
top quark condensate in a way consistent with the recent discovery
of a heavy top quark with mass about 180 GeV \cite{kn:TEVATRON}. 

We start with basic concepts of DSB with particular
emphasis on its
characterization in the high energy behavior through anomalous dimension 
(Section 2).
We then proceed to a general idea of TC and 
the role of anomalous dimension (Section 3).
  Detailed explanation will be given to how the large anomalous dimension
  arises in explicit dynamics, NJL model and gauged NJL model: 
In particular, the gauged NJL model
  is shown to be renormalizable
  in a non-perturbative sense, due to the large anomalous dimension and 
  the very presence of gauge interaction (Section 4, 5).
  We demonstrate that the entire coupling space of the gauged NJL model 
  encompasses a variety of tightly bound composite Higgs models, i.e.,
 walking technicolor, strong
  ETC technicolor and top quark condensate (Section 6). 
 In Section 7 detailed discussion will be given to the top quark condensate  
which actually predicted the top quark mass on the order of weak scale,
 exceptionally large compared with
all other quarks and leptons. We give a detailed comparison between
the original formulation of Miransky-Tanabashi-Yamawaki 
(MTY)\cite{kn:MTY89a,kn:MTY89b} and another one of
Bardeen-Hill-Lindner (BHL)\cite{kn:BHL90}. We then discuss
the renormalizability of the gauged NJL model which may be applied to
a possible modification of the top quark condensate (``top mode
walking GUT'').\cite{kn:ITY96}

\section{Basic Concepts in Dynamical Symmetry Breaking}

\subsection{Spontaneous Symmetry Breaking}

Here we first summarize basic concepts of the spontaneous symmetry breaking in
a way suitable to discussions on the DSB.

\subsubsection{Symmetry Realizations}
Let $G$ be a symmetry group of transformations with continuous 
parameters in such a way that
the action $S=\int d^4x 
{\cal L} (\phi,\partial_{\mu}\phi)$ is invariant under the
transformation $G$, 
$\delta \phi_i =\epsilon_A \delta^A \phi_i = -i\epsilon_A (T^A)_i^j\phi_j$,
where $T^A$ ($A=1,2,..., {\rm dim} G$) are the matrix representation of the generators
of $G$, i.e., ${\rm exp}(-i\epsilon_A T^A)\in G$. Then there exist conserved Noether currents
$J_{\mu}^A (x)$ corresponding to this symmetry, $\delta S=0 \Leftrightarrow
\partial^{\mu} J_{\mu}^A (x) =0$. We
may define (at least formally) conserved charge operators $Q^A\equiv \int d^3 x J_0^A (x)$, $\dot Q^A=0$,
which, if well-defined, become generators of the symmetry group $G$: 
\begin{equation}
\phi(x)\rightarrow
\phi'(x)=e^{i\epsilon_A Q^A}\phi(x) e^{-i\epsilon_A Q^A}, 
\label{transf}
\end{equation}
or
\begin{equation}
\delta^A \phi(x) = [iQ^A, \phi(x)] .
\end{equation}

Let us look at n-point Green functions
$
G_n (x_1,x_2,\cdot\cdot\cdot,x_n) \equiv \bra{0}T \phi_1(x_1)\phi_2(x_2)\cdot \cdot \cdot\phi_n(x_n)\ket{0},
$
where $T$ stands for the usual time-ordered product. 
They have all the information of quantum field theory, i.e., field operators and the vacuum.
Under the transformation of the above symmetry group $G$, $G_n(x_1,x_2,
\cdot\cdot\cdot,x_n)$ transforms as
\begin{equation}
G_n \rightarrow G_n'= \bra{0}T \phi_1'(x_1)\phi_2'(x_2)\cdot \cdot \cdot\phi_n'(x_n)\ket{0}
                    = \bra{0'}T \phi_1(x_1)\phi_2(x_2)\cdot \cdot \cdot\phi_n(x_n)\ket{0'},
\label{G_n'}
\end{equation}
where (\ref{transf}) and  $\ket{0'}=e^{-i\epsilon_A Q^A}\ket{0}$ were used. Our principal
interest lies in the variation of $G_n$, $\delta G_n\equiv G_n'-G_n =\epsilon_A \delta^A G_n$.
 
Now, there are two types of realizations (phases) of the symmetry:\\

(i) Wigner realization.\\

If a symmetry of the action is also the symmetry of the vacuum:
\begin{equation}
\ket{0}\rightarrow \ket{0'}=e^{-i\epsilon_A Q^A}\ket{0}=\ket{0},
\qquad (Q^A\ket{0}=0),
\end{equation}
then  we have $G_n' - G_n=0$ from (\ref{G_n'}), i.e.,
\begin{equation}
\delta^A G_n  =0
\end{equation}
for {\it all} Green functions. Namely, $Q^A\ket{0}=0 \Longrightarrow \delta^A \forall G_n=0$. 
Hence the symmetry is also the symmetry of physical
states in the Fock space constructed upon this vacuum: They are classified
according to the usual representation theory of the group.\\ 

(ii) Nambu-Goldstone (NG) realization.\\

Conversely, if 
a symmetry of the action is not the symmetry of the physical states in such a way that 
there exists {\it at least one} (not necessarily all) Green function such that
\begin{equation}
\delta^A G_n\ne 0,
\end{equation}
then the vacuum does not respect the symmetry; Symbolically,
\begin{equation}
\ket{0'}=e^{i\epsilon_A Q^A}\ket{0}\ne \ket{0} \qquad 
\left(Q^A\ket{0}\ne 0\right).
\end{equation}
Namely, $\delta^A \exists G_n \ne 0$$\Longrightarrow  Q^A\ket{0}\ne 0$.
We say that the symmetry is {\it spontaneously broken}.\\
 
The variation $\delta^A G_n$ is called an {\it order parameter} which in fact discriminates between the two realizations, 
Wigner (or disordered) phase 
with $\delta^A G_n= 0$ and NG (or ordered) phase with $\delta^A G_n\ne 0$.
We shall call it a ``{\it composite order parameter}''
if $n \geq 2$, and an ``{\it elementary order parameter}'' if
$n=1$.  Composite order parameters are in general 
{\it nonlocal} but could be local if we put $x_1=\cdot\cdot\cdot=x_n$.
Local composite order parameters are often 
called {\it condensates}. 
From (\ref{G_n'}) we may write
\begin{eqnarray}
\delta^A G_n (x_1,...,x_n) 
&=&
\bra{0}
[i Q^A, T \phi_1(x_1)\phi_2(x_2)\cdot \cdot \cdot\phi_n(x_n)] 
\ket{0}\nonumber \\
&=&
\bra{0}
T[i Q^A, \phi_1(x_1)]\phi_2(x_2)\cdot \cdot \cdot\phi_n(x_n) 
\ket{0}\nonumber \\
&&+\bra{0}
T\phi_1(x_1)[iQ^A, \phi_2(x_2)]\cdot \cdot \cdot\phi_n(x_n) \ket{0}
             +\cdot \cdot \cdot\nonumber \\
&&+\bra{0}T\phi_1(x_1) \phi_2(x_2)\cdot \cdot \cdot[iQ^A,\phi_n(x_n)] \ket{0}.
\label{order}
\end{eqnarray}
Note that the commutator $[i Q^A, \phi(x)]\equiv 
i\int d^4 z [J_0^A(z),\phi(x)]\delta(z^0-x^0)$
 is always well-defined even when the charge $Q^A$
itself is not. 

\subsubsection{Nambu-Goldstone Theorem}

Now we come to the basic theorem of the spontaneous symmetry breaking, namely the NG theorem:
If the continuous symmetry is spontaneously broken such that $\delta^A \exists G_n\ne 0$, then there exist massless spinless bosons
(NG bosons) coupled to the currents $J_{\mu}^A(x)$.
The proof is as follows:\\

Define 
\begin{equation}
M^A_{\mu}(q,x_1,...,x_n)\equiv \int d^4 z e^{iqz}
\bra{0} T J^A_{\mu}(z)\phi_1(x_1)\cdot \cdot \cdot\phi_n(x_n)\ket{0},
\end{equation}
where $\phi_1, \phi_2, ..., \phi_n$ are fields appearing in the Lagrangian.
The current conservation $\partial^{\mu} J^A_{\mu}=0$ leads to the Ward-Takahashi (WT) identity:
\begin{eqnarray}
\lim_{q_{\mu}\rightarrow 0}q^{\mu} M^A_{\mu}
&=&
\lim_{q_{\mu}\rightarrow 0}\int d^4 z e^{iqz} (i\partial_z^{\mu})
\bra{0} T J^A_{\mu}(z)\phi_1(x_1)\cdot \cdot \cdot\phi_n(x_n)
\ket{0},\nonumber\\
&=&\bra{0}T[i Q^A, \phi_1(x_1)]\phi_2(x_2)\cdot \cdot \cdot\phi_n(x_n) \ket{0}
             +\cdot \cdot \cdot\nonumber\\
&&+\bra{0}T\phi_1(x_1) \phi_2(x_2)\cdot \cdot \cdot[iQ^A,\phi_n(x_n)] 
\ket{0}\nonumber\\
&=&\delta^A G_n (x_1,...,x_n),
\end{eqnarray}
where use has been made of $\partial_z^{\mu} TJ^A_{\mu}(z)\phi(x)=[J^A_0, \phi(x)]\delta(z^0-x^0)$
and (\ref{order}).
Thus, if there exists at least one Green function $G_n$ such that $\delta^A G_n\ne 0$, then 
the corresponding $M^A_{\mu}$ must have a pole singularity at $q^2 =0$:
\begin{equation}
M^A_{\mu}(x_1,...,x_n) \sim \frac{q_{\mu}}{q^2}\delta^A G(x_1,...,x_n),
\label{NGpole}
\end{equation}
where the Lorentz index carried by $q_{\mu}$ implies a spinless particle.
The order parameter $\delta^A G_n\ne 0$ is nothing but a residue of the NG boson
pole.
This establishes existence of a massless spinless boson (NG boson) coupled to the current  $J^A_{\mu}$ (broken current) 
with strength $\delta^A G_n$. 

Generally, the generators of $G$, $\{T^A\}$, can be divided into two parts\\ 

(i) Unbroken ones $\{S^{\alpha}\}$ with $\delta^{\alpha} G_n=0$,
which span a subgroup $H$ ($H\subset G$),\\

(ii) Broken ones $\{X^a\}$ with $\delta^a G_n\ne 0$: Namely,
\begin{equation}
\{T^A\} = \{S^{\alpha}\in {\cal H}, X^a\in {\cal  G} -{\cal H}\},
\nonumber
\end{equation}
where ${\cal G}$ and ${\cal H}$ denote the algebra of $G$ and $H$, respectively.
In such a case $G$ is spontaneously broken down to a subgroup $H$.
As is obvious from the above  NG theorem,
there is {\it a one to one correspondence}
between the broken current $J^a_{\mu}$ (broken generator $X^a$) 
and the NG boson pole in $M^a_{\mu}$. 
Thus the number of independent NG bosons is given by that of independent
broken generators, ${\rm dim} G -{\rm dim} H$.

\subsection{Elementary versus Composite Order Parameters}

Let us now look at the pole residue of the NG boson appearing in the
current matrix $M^a_{\mu}(q,x_1,...,x_n)$:
\begin{equation}
M^a_{\mu}(q,x_1,...,x_n)\sim 
\bra{0}J_{\mu}^a(0)\ket{\pi^b(q)} 
\frac{i}{q^2} 
\bra{\pi^b(q)}T\phi_1(x_1)\cdot\cdot\cdot\phi_n(x_n)\ket{0}.
\end{equation}
Comparing this with (\ref{NGpole}),  we may write the order parameter
$\delta^a G_n$ as 
\begin{equation}
\delta^a G_n(x_1,...,x_n)=f_{\pi}\cdot
\bra{\pi^a(q_{\mu}=0)}T\phi_1(x_1)\cdot\cdot\cdot\phi_n(x_n)\ket{0},
\label{WT-BS}
\end{equation}
where $f_{\pi}$ is the ``decay constant'' of the NG boson defined by
\begin{equation}
\bra{0}J^a_{\mu}(x)\ket{\pi^b(q)}=
-i\delta^{ab}f_{\pi}q_{\mu}e^{-iqx}.
\label{fpi}
\end{equation}
Now the n-point order parameter $\delta^a G_n$ is traded for
the generic order parameter $f_{\pi}$ multiplied by 
the Bethe-Salpeter (BS) amplitude
$\bra{\pi^a(q)}T\phi_1(x_1)\cdot\cdot\cdot\phi_n(x_n)\ket{0}$
which plays a role of ``wave function'' of the bound state $\pi$.
Eq.(\ref{WT-BS}) is another expression of WT identity or the NG theorem
and is a basic formula for the soft NG boson emission vertex
(low energy theorem). We may distinguish
between the following two cases.\\  

Elementary Order Parameters:\\

The simplest order parameter is of course ``one-body'' order parameter
(elementary order parameter), in which case (\ref{WT-BS}) reads:
\begin{equation}
\bra{0}\delta^a \phi^b\ket{0}=\delta^{ab} f_{\pi}Z_{\pi}^{1/2}
\label{one-body}
\end{equation}
where the elementary field $\phi$ becomes an interpolating
field of the NG boson with  the renormalization constant  
$Z_{\pi}$ such that $\bra{\pi^a}\phi^b\ket{0}=\delta^{ab} Z_{\pi}^{1/2}$.
The linear $\sigma$ model and the Higgs Lagrangian in the standard model belong
to this category.\\
 
Composite order Parameters:\\

  On the other hand, if there exists no elementary order parameters
  $\bra{0} \delta^a \forall \phi^b\ket{0}=0$ but $f_{\pi}\ne 0$,
  then there must exist composite order parameters 
  $\delta^a \exists G_n \ne 0$ for $n\geq 2$
  which satisfies (\ref{WT-BS}). 
The NG boson $\pi$ is now a composite particle with non-zero BS amplitude
$\bra{\pi^a(q)}T\phi_1(x_1)
\cdot\cdot\cdot\phi_n(x_n)\ket{0}\ne 0$.
In such a case there also exists a {\it local}
 composite order parameter ({\it condensate}) which satisfies a relation 
similar to (\ref{WT-BS}):
\begin{equation}
\bra{0} \delta^a  (\phi_1(0)\phi_2(0)\cdot \cdot \cdot\phi_n(0))\ket{0}
=f_{\pi} \bra{\pi^a} \left(
\phi_1(0)\phi_2(0)\cdot \cdot \cdot\phi_n(0)\right)\ket{0}
\ne 0.
\label{condensate}
\end{equation}
This implies that the local composite operator
$
\left(\phi_1(0)\phi_2(0)\cdot \cdot \cdot\phi_n(0)\right)
$
has an overlapping with the NG boson $\pi$ and becomes its interpolating
field.
We call this case {\it dynamical symmetry breaking} (DSB).\footnote{
This definition is somewhat ambiguous when the system contains elementary
scalars. Even if the VEV of scalar is zero at tree level, quantum 
effects may give rise to a non-zero VEV as in the
  Coleman-Weinberg mechanism\cite{kn:CW}. Similar phenomenon
also takes place in the strongly
coupled Yukawa model, which very much resembles the DSB 
in the four-fermion (NJL) model.\cite{kn:KTY90,kn:KSTY94}
}
QCD is a good example of this category.
Note that when the elementary order parameter already exists,
then usually the composite order parameters are also  
non-zero, while the converse is not true.
Eqs.(\ref{WT-BS}) and (\ref{condensate}) are basic tools of DSB.

\subsection{$\sigma$ Model}

As a well-known example having an elementary order parameter, 
let us recall the linear $\sigma$ model. The Lagrangian is given by
\begin{eqnarray}
{\cal L}
&=&\bar\psi(i\xbar{\partial})\psi
-g_{NN\pi} \bar \psi (\sigma +i\gamma_5\tau^a\pi^a)\psi
+\frac{1}{2}\left((\partial_{\mu}\sigma)^2 
+(\partial_{\mu}\pi^a)^2\right) -V(\pi, \sigma),
\nonumber\\ 
V(\pi, \sigma) 
&=& \frac{\lambda_4}{4}\left((\sigma)^2+(\pi^a)^2-v^2\right)^2,
\label{sigmamodel}
\end{eqnarray}
where $\tau^a (a=1,2,3)$ are Pauli matrices in isospin space
of the nucleon doublet field $\psi=\left(
\begin{array}{c}
p\\
n
\end{array}
\right),
$
with $g_{NN\pi}$ and $\lambda_4$
being the Yukawa coupling of the nucleon and the quartic coupling of the 
mesons, respectively. The meson fields $\pi^a(x)$ and 
$\sigma(x)$ stand for  
the pion (in the massless limit) and the $\sigma$ meson 
(not existing in the Particle
Data Group Table, though often regarded as a very broad 
resonance), respectively.

 The Lagrangian can be cast into
the form:
\begin{eqnarray}
{\cal L}&=&\bar \psi_L (i\xbar{\partial})\psi_L +(L\rightarrow R)
-\sqrt{2} g_{NN\pi}\left(\bar \psi_L M\psi_R 
+\bar \psi_R M^\dagger \psi_L\right)\nonumber
\\
&+& \frac{1}{2}Tr(\partial_{\mu}M \partial^{\mu}M^\dagger)
+\frac{\lambda_4}{4}\left(Tr(M M^\dagger)-v^2\right)^2,
\label{M-sigma}
\end{eqnarray}
\begin{equation}
\psi_{L,R}\equiv \frac{1\mp \gamma_5}{2}\psi,\qquad
M\equiv \frac{1}{\sqrt{2}}(\sigma+i\pi^a \tau^a). 
\label{Mdef}
\end{equation}
It is
easy to see that this Lagrangian 
has a global symmetry under the 
$SU(2)_L\times SU(2)_R$ transformation:
\begin{eqnarray}
\psi_{L,R}(x)
&\rightarrow& 
\psi_{L,R}'(x)=
g_{L,R}\, \psi_{L,R}(x), 
\label{chiralpsi}\\
M(x)
&\rightarrow&
M'(x) =
g_L\, M(x)\, g_R^\dagger,
\label{chiralM}
\end{eqnarray}
where $g_{L,R}=e^{
-i\epsilon_{L,R}^a\frac{\tau^a}{2}
}$ $ \in SU(2)_{L,R}$, or
$\delta_{L,R}^a \psi_{L,R}=i[Q_{L,R}^a,\psi_{L,R}]$ 
$=-i\frac{\tau^a}{2} \psi_{L,R}$.
For $\epsilon_L^a=\epsilon_R^a=\epsilon^a$ we have a vector transformation
\begin{eqnarray}
\delta^a \psi &=&[iQ^a,\psi]=i[Q^a_R+Q^a_L,\psi]=-i\frac{\tau^a}{2} \psi,\\
\delta^a M&=&
-i[\frac{\tau^a}{2},M]=\frac{i}{\sqrt{2}}\epsilon^{abc}\pi^b\tau^c,
\end{eqnarray} 
which forms the usual isospin $SU(2)_V$ symmetry.
On the other hand, for 
$-\epsilon_L^a=\epsilon_R^a=\epsilon_5^a$ we have an axialvector transformation \begin{equation}
\delta_5^a \psi=[iQ^a_5,\psi]
=i[Q^a_R-Q^a_L,\psi]=-i\gamma_5\frac{\tau^a}{2} \psi,
\label{axial-psi}
\end{equation}
($\delta_5^a \bar \psi=-\bar \psi i\gamma_5\frac{\tau^a}{2}$).
As to the axialvector transformation of $M$, we have
\begin{equation}
\delta_5^a M=
i\{\frac{\tau^a}{2},M\}=\frac{1}{\sqrt{2}}(i\delta^{ab}\tau^b \sigma - \pi^a),
\end{equation}
or
\begin{equation}
\delta_5^a \pi^b=[iQ_5^a,\pi^b]=\delta^{ab}\sigma, \qquad
\delta_5^a\sigma = [iQ_5^a, \sigma] = -\pi^a.
\end{equation}

Suppose that 
the vacuum is chosen as
\begin{equation}
 \VEV{\delta_5^a\pi^b}=\VEV{[iQ_5^a,\pi^b]}=\delta^{ab}\VEV{\sigma}\ne 0,
\qquad  \VEV{\delta_5^a\sigma} =\VEV{[iQ_5^a, \sigma]} = -\VEV{\pi^a}=0,
\label{sigmaVEV}
\end{equation}
then the chiral symmetry $SU(2)_L\times SU(2)_R$ is spontaneously broken down
to $SU(2)_V$ through the {\it elementary order parameter}. 
Recalling (\ref{one-body}), one finds
$f_{\pi}Z_{\pi}^{1/2}=\VEV{\sigma}$. 

 Actually at tree level, the wine-bottle potential 
 in (\ref{sigmamodel}) has infinitely 
 degenerate ground states at $\VEV{\sigma^2+(\pi^a)^2} =v^2$, among which
 we can always choose (through appropriate $SU(2)_L \times SU(2)_R$ rotation) 
 the unique vacuum satisfying (\ref{sigmaVEV}), i.e.,
$f_{\pi}=\VEV{\sigma}=v \ne 0$ ($Z_{\pi}=1$ at level) 
and $\VEV{\pi^a}=0$. 
Then the spectrum of the theory can readily be 
read off from the Lagrangian at this vacuum. The curvature
of the potential $V(\pi, \sigma)$ in (\ref{sigmamodel}) yields the 
$({\rm mass})^2$:
\begin{eqnarray}
 m_{\sigma}^2 
&=&
 \frac{{\partial}^2 V}{\partial \sigma \partial \sigma}
 \Bigg|_{\VEV{\sigma}=v,\VEV{\pi} =0} =2\lambda_4 v^2 >0,\nonumber \\
 m_{\pi}^2 
&=&
\frac{{\partial}^2 V}{\partial \pi \partial \pi}
\Bigg|_{\VEV{\sigma}=v,\VEV{\pi} =0} = 0,
\end{eqnarray}
where the flat curvature in $\pi^a$ directions corresponds to the fact that
$\pi^a$ are massless NG bosons.  

On this spontaneous chiral symmetry breaking (\SxSB) 
vacuum the nucleon also acquires a mass through the Yukawa coupling:
\begin{equation}
-g_{NN\pi} \bar \psi  \left(\sigma +i\gamma_5\tau^a\pi^a\right)\psi
\Bigg|_{\VEV{\sigma}=v,\VEV{\pi} =0}
=-\left(g_{NN\pi}\VEV{\sigma}\right) \bar \psi \psi
-g_{NN\pi} \bar \psi \left(\sigma' +i\gamma_5\tau^a\pi^a\right)\psi,
\label{N-mass}
\end{equation}
where $\sigma'\equiv \sigma - \VEV{\sigma}$. The first term
indeed behaves like a mass term, with a  mass given by 
\begin{equation}
m_N =g_{NN\pi}\VEV{\sigma}=g_{NN\pi} v=g_{NN\pi}f_{\pi}
\label{GT}
\end{equation} 
at tree level. This is the Goldberger-Treiman relation (with $g_A=1$
because of tree-level).
Now, it is evident that the various order parameters 
$m_N$, $\VEV{\sigma}$, $f_{\pi}$ are
proportional to each other.

Through the above popular arguments {\it at Lagrangian level}, 
one might have gotten an impression as if the
symmetry were broken already at Lagrangian (operator) level. 
Indeed the mass term $m_N\bar \psi \psi$ as it stands is not
invariant under the chiral symmetry (\ref{chiralpsi}), 
since $\VEV{\sigma}$ is now
just a constant and no longer transform to cancel the transformation of
$\bar \psi\psi$. However, in the case of the spontaneous symmetry breaking 
the symmetry is {\it not} broken at operator (Lagrangian)
level, but solely on the vacuum. The Lagrangian must keep the invariance at any
change of variables. So, what happened to the nucleon mass?
Does it really break the Lagrangian symmetry? The answer is of course no.

Actually, the nucleon field $\psi$ in (\ref{N-mass}) is 
not an appropriate field
to describe this phenomenon properly at Lagrangian level. 
We shall explain this in terms of the nonlinear realization
(For a review see Ref.26).
First note that any complex matrix 
$M\equiv (\sigma+i\pi^a \tau^a)/\sqrt{2}$
can be written as 
$M= \tilde H U=\xi_L^\dagger H\xi_R$, where $\tilde H 
(\equiv \xi_L^\dagger H \xi_L)$ is a Hermitian matrix,
and $U(\equiv \xi_L^\dagger \xi_R)$ and $\xi_L^\dagger=\xi_R= \xi$ are
unitary matrices (polar decomposition). The Hermite matrix (radial mode)
$H$ can always be 
diagonalized as $H=\frac{1}{\sqrt{2}}
\left(
\begin{array}{cc}
\hat \sigma &0\\
0  &\hat \sigma
\end{array}
\right)
$
such that $\VEV{\hat \sigma} =\VEV{\sigma}=v$, 
while the unitary matrix (phase modes)
may be parameterized in terms of the NG bosons $\hat \pi$:
$\xi=e^{i\hat\pi^a(\frac{\tau^a}{2}) /v}$,
where $\hat \pi \sim \pi$ and $\hat \sigma \sim \sigma$. 
(This polar decomposition
makes sense only when $\VEV{H}\ne 0$, i.e., 
the chiral symmetry is spontaneously
broken.) 
The new fields $\xi_{L,R}$ and $H$ transform as 
$
\xi_{L,R}\rightarrow h\, \xi_{L,R}\, g_{L,R}^\dagger,
$
$
H \rightarrow  h\, H\, h^\dagger,
$
where $h \in SU(2)_V$ and $g_{L,R} \in SU(2)_{L,R}$.
Then we can introduce an appropriate ``nucleon'' field 
$\Psi_{L,R} \equiv \xi_{L,R}\,\psi_{L,R}$, 
which transform as  
$
\Psi_{L,R} \rightarrow h \Psi_{L,R}.
$
Now the Yukawa term reads:
\begin{equation}
\bar \psi_L M\psi_R= \bar \Psi_L H \Psi_R
=\bar \Psi_L \VEV{H} \Psi_R
+\bar \Psi_L H' \Psi_R,
\label{invmass}
\end{equation}
and ($L \leftrightarrow R$),
where $H'\equiv H -\VEV{H}$. The first term of R.H.S. in (\ref{invmass}) 
yields precisely the same 
nucleon mass as before but this time it is {\it invariant} under the 
chiral symmetry, since $\Psi_{L,R}$ now transforms only under $SU(2)_V$ but
not $SU(2)_L\times SU(2)_R$. Note that 
$\psi_{L,R}=e^{\mp i\hat\pi^a(\frac{\tau^a}{2}) /v }
\Psi_{L,R}=\Psi_{L,R}+\cdot\cdot\cdot$, so that the above popular arguments
at Lagrangian level are  effectively still correct.  

Thus the arguments at Lagrangian level are rather tricky. Here we return to
rather straightforward arguments based on 
the Green functions which contain information on 
both the operator and the vacuum. Non-invariance of the Green functions are
solely due to the vacuum structure but not to the operator, as we discussed
in the beginning of this lecture.
Let us look at (Fourier transform of) 2-point function of the nucleon, 
the full propagator 
$G_2(p)=S(p)={\cal FT}
\bra{0} T\left( \psi(x)\bar\psi(0) \right) \ket{0},
$
whose variation is a {\it composite order parameter}
(see (\ref{order}) and (\ref{WT-BS})):
\begin{equation}
\delta_5^a S(p)=
{\cal FT} \bra{0} T\left( [iQ_5,\psi(x)\bar\psi(0)] \right) \ket{0}
=f_{\pi}\cdot 
{\cal FT} \bra{\pi^a} T\left( \psi(x)\bar\psi(0) \right) \ket{0}
\ne 0.
\label{S}
\end{equation}
From (\ref{axial-psi}) we can rewrite the L.H.S. as
\begin{equation}
\delta_5^a S=
S\,\{-i\gamma_5\frac{\tau^a}{2}, S^{-1} \}\,S 
=S\, \left(\gamma_5 \tau^a Z_{\psi}^{-1}(-p^2) \Sigma(-p^2) \right)\, S,
\end{equation}
where $i S^{-1}(p)=Z_{\psi}^{-1}(-p^2) \left(\xbar{p} -\Sigma(-p^2)\right)$ 
(in space-like region $p^2<0$), with 
$Z_{\psi}(-p^2)$ and
$\Sigma(-p^2)$ being the wave function renormalization 
and mass function of the
nucleon, respectively.
This implies that non-invariance of the Green function or
the vacuum symmetry breaking is signalled by the appearance 
of mass function $\Sigma(-p^2)$ of the nucleon 
which is massless at Lagrangian level: $\delta_5^a S(p)\ne 0\Leftrightarrow
\Sigma(-p^2)\ne 0$.

Now define the amputated BS amplitude 
\begin{equation}
\tilde \chi_{\pi}^a(p,q)\equiv S^{-1}(p+q)\,
{\cal FT} \bra{\pi^a} T\left( \psi(x)\bar\psi(0) \right) \ket{0}\,
S^{-1}(p),
\end{equation}
then (\ref{S}) reads
\begin{equation}
 \gamma_5 \tau^a  \Sigma(-p^2) 
 = f_{\pi}\cdot \chi_{\pi}^a(p,0) \ne 0,
\label{WT-Mass}
\end{equation}
where $\chi_{\pi}^a(p,0)\equiv Z_{\psi}\tilde\chi_{\pi}^a(p,0)$ 
is a renormalized amputated BS amplitude.
At tree level we have $f_{\pi} =\VEV{\sigma}$
and $\chi_{\pi}^a(p,0) =\gamma_5 \tau^a g_{NN\pi}={\rm const.}$,
then (\ref{WT-Mass}) implies a constant nucleon mass
$\Sigma(-p^2)\equiv m_N=f_{\pi}g_{NN\pi}=g_{NN\pi}\VEV{\sigma}$,
which is nothing but the Goldberger-Treiman relation (\ref{GT})
obtained through the Lagrangian level arguments.

Note that the BS amplitude $\chi_{\pi}^a(p,0)\ne 0$ in this case 
just means an already existing Yukawa vertex among the elementary fields 
in the Lagrangian and hence never implies that $\pi$ is a ``composite''
of $N \bar N$ (at least at tree level). 
 This reflects the fact that main origin
of the symmetry breaking $f_{\pi} \ne 0$ is due to
the elementary order parameter $\VEV{\delta_5^a \pi^b}
=\delta^{ab}\VEV{\sigma}\ne 0$
but not due to the composite one (fermion pair condensate $\VEV{\bar \psi \psi}$) related to the compositeness of $\pi$.\footnote{
This statement is no longer valid if the Yukawa coupling is strong
enough to trigger the symmetry breaking.\cite{kn:KTY90,kn:KSTY94} 
} 
This model is a good example to show that if there exists
 an elementary order parameter, then  
 composite (nonlocal) order parameters also exist in general.
However, the latter are only secondary objects in this model: For example, 
even when $f_{\pi}=\VEV{\sigma}\ne0$, $\Sigma(-p^2)$ could be zero
(if $g_{NN\pi}=0$).

\subsection{QCD}

\subsubsection{Chiral Symmetry}

Now we discuss QCD where no elementary order parameters exist while
composite ones actually do. For simplicity we confine ourselves 
to the 2-flavor quarks 
$
q=\left(
\begin{array}{c}
u\\
d
\end{array}
\right)
$ 
whose masses are very small
compared with the QCD scale $\Lambda_{QCD}$, i.e.,  $m_u, m_d \ll \Lambda_{QCD}$.
These masses, called {\it current masses}, are entirely due to the Higgs VEV
through the Yukawa coupling in the Glashow-Salam-Weinberg model and have
nothing to do with the QCD dynamics.

The QCD Lagrangian is given by
\begin{equation}
{\cal L}=\bar q (i\xbar{D} -{\cal M})q-\frac{1}{4}F_{\mu\nu}^{\alpha}
F_{\alpha}^{\mu\nu}+({\rm gauge\: fixing})+({\rm FP\: ghost}), 
\end{equation}
where $D_{\mu}=\delta^{ij}
(\partial_{\mu} - g \frac{\lambda^{\alpha}}{2}A^{\alpha}_{\mu})$
($\alpha=1,\cdot\cdot,8$)
is a unit matrix in flavor space $i,j =(u,d)$, 
with $g$ being the gauge coupling,
and $F_{\mu\nu}^{\alpha}$ is a field strength of the gluon field
$A_{\mu}^{\alpha}$, and 
the quark (current) mass matrix ${\cal M}$ is 
diagonal with eigenvalues $(m_u, m_d)$. 
In the limit ${\cal M}\rightarrow 0$ the Lagrangian 
possesses a chiral $SU(2)_L\times SU(2)_R$ symmetry under
the transformation:
\begin{equation}
q_{L,R}(x) \rightarrow 
q_{L,R}' (x)=e^{-i \epsilon_{L,R}^{a}\frac{\tau^a}{2}}\:q_{L,R}(x),
\end{equation}
which is expected to be spontaneously broken down to $SU(2)_V$. 
The broken current is
the axialvector current:
\begin{equation}
J_{5 \mu}^a =\bar \psi i\gamma_5\frac{\tau^a}{2}\psi.
\label{axialvector}
\end{equation}

\subsubsection{Composite Order Parameters in QCD}

In contrast to the $\sigma$ model, QCD has no elementary
order parameters. If the quark and gluon fields were order parameters,
then the Lorentz invariance, color symmetry and charge symmetry 
would have been spontaneously broken in QCD in contrast to the reality.
Then only possible order parameters are
composite ones, variation of n-point Green functions or that of 
local composite fields. 
A relevant composite order parameter (2-point function)
takes the same form as (\ref{S}):
\begin{equation}
\delta_5^a S(p)=
{\cal FT} \bra{0} T\left( [iQ_5, q(x)\bar q(0)] \right) \ket{0}
=f_{\pi}\cdot 
{\cal FT} \bra{\pi^a} T\left( q(x)\bar q(0) \right) \ket{0}
\ne 0,
\label{Sq}
\end{equation}
which then leads to the same relation as (\ref{WT-Mass}):
\begin{equation}
 \gamma_5 \tau^a  \Sigma(-p^2) 
 = f_{\pi}\cdot \chi_{\pi}^a(p,0) \ne 0,
\label{WT-QMass}
\end{equation}
where $\Sigma(-p^2)$, now a {\it dynamical mass} of quark,
signals the spontaneous chiral symmetry breaking
due to the QCD dynamics. We may define an ``on-shell'' dynamical mass
$m^*$ as $\Sigma ({m^*}^2)=m^*$, which is often called {\it constituent mass}
(it also includes the effects of the explicit breaking due to the current 
mass).  In contrast to the $\sigma$ model where 
$\chi_{\pi}^a(p,0)=g_{NN\pi}\tau^a\gamma_5$ (tree level), there is no
Yukawa coupling $g_{qq\pi}$ at Lagrangian level in QCD.
However, we have an ``induced'' Yukawa vertex $\chi_{\pi}^a(p,0)$ which is
a ``wave function'' of $\pi$ as a composite of $q\bar q$ and is 
related to the dynamical mass $\Sigma(-p^2)$ through (\ref{WT-QMass}).

According to (\ref{condensate}),
we may also consider a two-body {\it local} composite order parameter
corresponding to (\ref{Sq}): 
\begin{equation}
\VEV{[i Q_5^a, \bar q i\gamma_5\tau^b q]}
=f_{\pi} \cdot \bra{\pi^a}\bar q i\gamma_5\tau^b q\ket{0}
= \delta^{ab} \VEV{\bar q q}\ne 0,
\end{equation}
which implies that the composite operator $\bar q i\gamma_5\tau^a q$ is an 
interpolating field of the composite NG boson, the pion $\pi^a$.
Note that $\bar q i\gamma_5\tau^a q$ and $\bar q q$ 
transform in the same way as  $\pi^a$ and $\sigma$ 
in the $\sigma$ model and actually are  interpolating fields of
the pion and the $\sigma$ meson, respectively: $\pi^a \sim
\bar q i\gamma_5\tau^a q$, $\sigma \sim \bar q q$. 

Thus the composite order parameters $\Sigma(-p^2)$ and $\VEV{\bar q q}$
in QCD necessarily imply compositeness of the pion, $\chi_{\pi}^a(p,0)\ne 0$
and $\bra{\pi^a}\bar q i\gamma_5\tau^a q\ket{0}\ne 0$, respectively.

\subsubsection{High Energy Behavior of Composite Order Parameters}

Now we are interested in the high energy (short distance) behavior of such
composite order parameters, which can probe the underlying dynamics
relevant to the composite NG boson.
Actually, detailed information of the underlying dynamics is reflected on
the behavior of nonlocal order parameter $\Sigma(-p^2)$ at
$-p^2 \gg \Lambda_{QCD}^2$, or equivalently that of local order
parameter (condensate) $\bra{0}(\bar q q)_{\Lambda}\ket{0}$ at 
$\Lambda\gg \Lambda_{QCD}$,
where $\Lambda$ is a high energy scale 
to renormalize the condensate. As a low
energy scale we take the scale parameter of QCD, $\Lambda_{QCD}$, which
is typically of order 100 MeV - 1 GeV and actually characterizes the 
scale of the order parameters $f_{\pi} \simeq 93 {\rm MeV}$,
$m^* \simeq 300 {\rm MeV}$ or $\VEV{\bar q q}_{\Lambda_{QCD}}
 \simeq -(250 {\rm MeV})^3$.

We first study a local composite order parameter, a condensate
$\VEV{\bar \psi \psi}$ for a generic fermion $\psi$. 
The renormalization of the condensate operator is 
\begin{equation}
\left(\bar \psi \psi\right)_0
\equiv 
\left(\bar \psi \psi\right)_{\Lambda}
=Z_m^{-1} \left(\bar \psi \psi \right)_{\mu},
\end{equation} 
where the suffix 0 stands for a bare quantity at UV cutoff $\Lambda$
($\gg \mu$, $\mu$: reference renormalization point)
and 
$Z_m=Z_m\left(\frac{\Lambda}{\mu}\right)$ 
is the renormalization 
constant of the current quark mass
\begin{equation}
m_0\equiv m_{\Lambda}=Z_m m_{\mu}, 
\label{mass-ren}
\end{equation}
so that the mass term is renormalized as  
 $m_0 (\bar \psi \psi)_0=m_{\mu}(\bar \psi \psi)_{\mu}$.
We introduce anomalous dimension  
\begin{equation}
\gamma_m(g_{\mu}) \equiv \mu \frac{\partial}{\partial \mu}\ln Z_m.
\end{equation}
This can be inverted into
\begin{equation}
Z_m =
 \exp \left[- \int_0^{t_{\Lambda}} \gamma_m(t') dt' \right],
\end{equation}
where we have defined $\gamma_m(t)\equiv \gamma_m(\bar g(t))$
in terms of a running coupling $\bar g(t)$ which satisfies
$\frac{d}{d t} \bar g(t)=\beta(\bar g(t))$ such that $\bar g(0)=g_{\mu}$,
with $\beta (g_{\mu}) \equiv  
\mu \frac{\partial}{\partial \mu}g_{\mu}$ and
$t\equiv \frac{1}{2}\ln\frac{-p^2}{\mu^2}$
($t_{\Lambda}\equiv 
 \frac{1}{2}\ln\frac{\Lambda^2}{\mu^2}$).
Thus the renormalization effects on the condensate at high energy scale 
 $\Lambda (\gg \mu=O(\Lambda_{QCD}))$ is governed by
 the anomalous dimension:
\begin{equation}
 \VEV{\bar \psi \psi}_\Lambda = Z_m^{-1} \VEV{\bar \psi \psi}_{\mu}\\
 = \exp \left[ \int_0^{t_\Lambda} \gamma_m(t') dt' \right]
   \cdot \VEV{\bar \psi \psi}_{\mu}. 
\label{condensate-ren}
\end{equation}
The positive anomalous dimension enhances the condensate at high
energy scale. The above result remains the same in the chiral symmetry limit
$m_{\mu}\rightarrow 0$.
 
The same enhancement factor due to the anomalous dimension also appears
in the nonlocal order parameter.
Using the Wilson's operator product expansion (OPE), 
we expand the product of operators in terms of a series of 
local composite operators at short distances:
\begin{equation}
T\left( \psi(x) \bar \psi(0) \right) \mathop{\sim}^{x \rightarrow 0}
C_{\bf 1}(x) {\bf 1} +
C_{\bar \psi \psi}(x) (\bar \psi \psi)_{\mu}+\cdot\cdot\cdot, 
\label{OPE}
\end{equation}
where each term is factorized into the $x$-dependent c-numbers
$C_i(x)$ called Wilson coefficients and the $x$-independent  
renormalized local composite operators $O_i$. The operators are placed  
in the order of increasing canonical mass dimension; 
{\bf 1} is a unit operator (dimension 0), 
$(\bar \psi \psi)_{\mu}$ is
a  quark condensate operator (dimension 3) renormalized at 
$\mu$ and $\cdot\cdot\cdot$ stands for the operators of 
higher dimensions. Since the L.H.S. has a definite mass dimension ($=3$
in this case), the coefficient corresponding to lower dimensional 
operator is expected to be 
more singular (more dominant) at $x\rightarrow 0$: $C_i (x) \sim
x^{-3+d_i}$ for the operator with dimension $d_i$. 
Taking Fourier transform of (\ref{OPE}), 
we have 
\begin{equation}
-i S(p) \mathop{\sim}^{p \rightarrow \infty}
 C_{\bf 1}(p)\VEV{{\bf 1}}+C_{\bar \psi \psi}(p) \VEV{\bar \psi \psi}_{\mu}
 +\cdot\cdot\cdot,
\label{OPE-Sp}
\end{equation}
which should be compared with the expansion of L.H.S.:
\begin{equation}
-i S(p)
=Z_{\psi}(-p^2)\frac{1}{\xbar{p}-\Sigma(-p^2)}
 \mathop{\simeq}^{p \rightarrow \infty}
Z_{\psi}(-p^2)\left(\frac{1}{\xbar{p}}+\frac{\Sigma(-p^2)}{p^2}+
\cdot\cdot\cdot\right),
 \label{RGE-S}
\end{equation}
where $Z_{\psi}(-p^2)$ is corresponding to
 a wave function renormalization of the fermion
field and $Z_{\psi}(-p^2) \rightarrow 1$ ($-p^2 \rightarrow \infty$)
in Landau gauge. (Hereafter we confine ourselves to the Landau gauge.) 
If the corresponding operators were scaling
according to the canonical dimension,
then $C_{\bf 1}(p)$ and $C_{\bar \psi \psi}(p)$ would behave 
as $C_{\bf 1}(p)\sim 1/\xbar{p}
$ 
and $C_{\bar \psi \psi}(p)
\sim 1/(p^2)^2$, respectively: Namely,
\begin{equation}
\Sigma(-p^2) \sim
\frac{\VEV{\bar \psi \psi}_{\mu}}{p^2}.
\end{equation}
 
However, 
the behavior of $C_{\bf 1}$ and $C_{\bar \psi \psi}$ 
will be modified by the renormalization effects through the 
anomalous dimension.
The Wilson coefficients satisfy 
the renormalization-group equation (RGE)  
whose solutions read:
\begin{eqnarray}
C_{\bf 1} (p)&=& 
 \exp \left[-\int_0^t 2\gamma_{\psi}(t') dt' \right]
\frac{1}{\xbar{p}}
+\cdot\cdot\cdot,
\nonumber\\
C_{\bar \psi \psi}(p) &=& c_{\bar \psi \psi}(p)
\frac{1}{(p^2)^2}\VEV{\bar \psi \psi}_{\mu}
 \exp \left[\int_0^t
  \gamma_m(t')-2\gamma_{\psi}(t') dt' \right]+\cdot\cdot\cdot,
\label{coefficients}
\end{eqnarray}
where $\gamma_{\psi}\equiv  \mu 
\frac{\partial}{\partial \mu}\ln Z_{\psi}^{1/2}$,
with $Z_{\psi}=Z_{\psi}\left(\Lambda/\mu\right)$ being the 
 fermion wave function renormalization constant,
 and $c_{\bar \psi \psi}(p)$ is
a part of the Wilson 
coefficients not determined by RGE alone
(determined by details of the dynamics).
Comparing both sides of (\ref{OPE-Sp}), i.e., 
(\ref{RGE-S}) and (\ref{coefficients}), we get
\begin{equation}
Z_{\psi}(-p^2) 
\mathop{\simeq}^{-p^2 \rightarrow \infty}
 \exp \left[-\int_0^t 2\gamma_{\psi}(t') dt' \right],
\end{equation}
\begin{equation}
\Sigma(-p^2)
\mathop{\simeq}^{-p^2 \rightarrow \infty}
\frac{\VEV{\bar \psi \psi}_{\mu}}{p^2}\cdot
c_{\bar \psi \psi}(p)\:
 \exp \left[\int_0^t \gamma_m(t') dt' \right].
\label{mass-asy}
\end{equation}
Note that 
order parameters are in general proportional to each other: 
The nonlocal order parameter $\Sigma (-p^2)$ is in 
fact proportional to the local one
$\VEV{\bar \psi \psi}_{\mu}$.
Moreover, the nonlocal order parameter $\Sigma(-p^2)$ has 
the same enhancement factor
\begin{equation}
 \exp \left[\int_0^t \gamma_m(t') dt' \right]
\end{equation}
as the local one 
 $\VEV{\bar q q}_{\Lambda}$ in (\ref{condensate-ren}),
 with a simple replacement $t_{\Lambda} \rightarrow t$.

\subsubsection{High Energy Behavior of Composite Order Parameters in QCD}

Now, the  QCD coupling is asymptotically free and small in
the high energy region.
The one-loop
beta function and the anomalous dimension are given by
$\beta (g) = -bg^3$ 
with $b = \frac{1}{(4\pi)^2}(11N_c -2 N_f)/3 $, 
and $\gamma_m = c g^2$ where $c=\frac{1}{(4\pi)^2}
 6 C_2 (\mbox{\boldmath$F$})$,
with $C_2(\mbox{\boldmath$F$})=(N_c^2-1)/(2N_c)$ being the quadratic Casimir of
the fermion representation $\mbox{\boldmath$F$}=\mbox{\boldmath$N_c$}$
$(N_c=3)$.
Then the  anomalous
dimension is logarithmically vanishing: 
\begin{equation}
\gamma_m (\bar g (t))=c {\bar g(t)}^2= \frac{A}{2\bar t}
\qquad (\bar t \equiv \frac{1}{2}\ln \frac{-p^2}
{{\Lambda_{QCD}}^2}=
t+\ln\frac{\mu}{\Lambda_{QCD}}),
\label{gammamQCD}
\end{equation}
with $A =c/b= 24/(33-2N_f)$.
This gives rise to (only) a logarithmic enhancement of the canonical result:
\begin{equation}
 \exp \left[ \int_0^t \gamma_m(t') dt' \right]
  \simeq \exp \left[\frac{A}{2} \ln (\frac{\bar t}{{\bar t}_\mu})\right]
  =    \left(\frac{\ln \frac{-p^2}{ {\Lambda_{QCD}}^2 }}
                  {\ln \frac{{\mu}^2}{ {\Lambda_{QCD}}^2 } }
      \right)^{A \over 2}.
\label{eq:(3)}
\end{equation}

More specifically, (\ref{condensate-ren}) reads
\begin{equation}
\VEV{\bar q q}_{\Lambda}=
    \left( \frac{ \ln \frac{\Lambda}{ \Lambda_{QCD} }
                 }
                 { \ln \frac{\mu}{ \Lambda_{QCD}}
                 }
    \right)^{A \over 2}
                 \VEV{\bar q q}_{\mu},
\label{QCDcond-ren}
\end{equation}
and (\ref{mass-asy}) reads\cite{kn:Lane74}
\begin{equation}
\Sigma(-p^2)
 \mathop{\simeq}^{-p^2 \rightarrow \infty}
\frac{1}{p^2}\:
\left[\frac{\frac{2\pi^2 A}{N_c} 
}
{ \left( \ln \frac{{\mu}^2}{ {\Lambda_{QCD}}^2 } \right)^{A \over 2} 
}\cdot\VEV{\bar q q}_{\mu}\right]\cdot
 \left( \ln \frac{-p^2}{ {\Lambda_{QCD}}^2 } 
                                  \right)^{{A \over 2} -1},
\label{QCDmass-asy}
\end{equation}
where the extra ${\rm (log)}^{-1}$ factor came from 
$c_{\bar \psi \psi}(-p^2)$
$=(1/4N_c)3C_2(\mbox{\boldmath$F$}) \bar g^2$
$=(\pi^2A/(N_c\bar t)$, which is due to
one-gluon exchange graph in Landau gauge, the lowest diagram
giving rise to the condensate operator in 
OPE.\cite{kn:Lane74,kn:Mira94}\footnote{
Only with this extra log factor, the mass function can correctly reproduce 
(\ref{QCDcond-ren}) through the condensate integral;
\begin{equation}
\VEV{\bar q q}= -TrS(p)=-\frac{N_c}{4\pi^2}
\int_0^{\Lambda^2} d x \frac{x \Sigma(x)}
{x+\Sigma(x)^2}, \qquad x\equiv -p^2>0.
\end{equation}
}
Thus the 
dynamical mass 
is rapidly damping in high energy, $1/p^2$,
roughly the result of canonical dimension arguments up to logarithm. Since
the QCD coupling is vanishingly small in high energy, quantum corrections 
due to vanishingly small anomalous dimension are accordingly  very small, only 
logarithmic  deviation. This implies that \SxSB
effects diminish rapidly when the coupling 
(attractive force) tends to zero in high energy.

\subsubsection{Pagels-Stokar Formula for $f_{\pi}$}

The generic order parameter $f_\pi$ defined by (\ref{fpi}) 
for the axialvector current (\ref{axialvector}) 
is the decay constant of the pion which may be calculated through graphical
consideration:
\begin{equation}
f_{\pi}q_{\mu}\delta^{ab}
=i\bra{0}J^a_{5\mu}(0)\ket{\pi^b(q)}=
-i \int \frac{d^4 p}{(2\pi)^4} Tr\left[ S(p+q) \chi_{\pi}^b(p,q) S(p) \gamma_{\mu}
\gamma_5 \frac{\tau^a}{2} \right],
\label{fpiinteg}
\end{equation}
where the amputated BS amplitude of the NG boson $\chi_{\pi}^a (p,q)$ 
may be determined by the BS equation.
Instead of solving the BS equation, however, 
here we use the famous Pagels-Stokar (PS)
 formula\cite{kn:PS79} which expresses the decay constants
in terms of dynamical mass function $\Sigma (-p^2)$ of the condensed 
fermion.
 Taking derivative of (\ref{fpiinteg}) with respect to $q^\nu$ and setting 
$q^{\nu}=0$,
we get
\begin{equation}
g_{\mu\nu}\delta^{ab}f_{\pi}
=\int \frac{d^4 p}{(2\pi)^4} Tr\left[\frac{\partial S(p)}{\partial p^\nu}
 \chi^b(p,0) S(p) \gamma_{\mu}
\gamma_5 \frac{\tau^a}{2} 
+S(p) \chi_{\pi}^b(p,0)' S(p) \gamma_{\mu}
\gamma_5 \frac{\tau^a}{2}
\right],
\end{equation}
where $\chi_{\pi}^a(p,q)'\equiv
(\partial/\partial p^{\nu})\chi_{\pi}^a(p,q)$. 

The PS formula is
obtained simply by ignoring the second term with the derivative $\chi_{\pi}'$. 
This is known to be a good approximation
when the ladder approximation is good.\cite{kn:ABKMN90}
Then we only have to evaluate the first term which is written {\it 
only in terms of
the mass function} $\Sigma(-p^2)$, since $\chi_{\pi} (p,0)$ is related to
  $\Sigma(-p^2)$ through the WT identity (\ref{WT-QMass}):
\begin{equation}
\chi_{\pi}^a(p,0)
=\frac{\gamma_5 \tau^a  \Sigma(-p^2)}{f_{\pi}}. 
\end{equation}
Thus we obtain the PS formula for the NG boson in the spontaneous 
symmetry breaking
$SU(2)_L \times SU(2)_R\rightarrow SU(2)_V$:
\begin{equation}
f_{\pi}^2 =\frac{N_c}{4\pi^2} \int_0^{\infty} dx\cdot x
\frac{\Sigma(x)^2-\frac{x}{4}\frac{d}{dx}\Sigma^2(x)}{(x+\Sigma^2(x))^2},
\label{PS-QCD}
\end{equation}
where $x\equiv -p^2$. 

We can easily see that $f_\pi$ also depends 
on the anomalous dimension or the damping rate of the mass function,
though its dependence is much milder than that of the condensate.
Although the rapid damping mass function  (\ref{QCDmass-asy}) trivially  
yields convergent integral for $f_{\pi}$ in the QCD case,
its convergence is  highly nontrivial in the general case 
and is actually related to the 
renormalizability/nontriviality of the 
gauged NJL model whose mass function is very slowly damping due
to large anomalous dimension as will be discussed in later sections.

\subsection{High Energy Theories and Anomalous Dimension}

If in contrast to QCD the theory has a non-vanishing anomalous dimension
$\gamma_m (t) \simeq \gamma_m \ne 0$ due to non-vanishing coupling 
constant (behaving as a nontrivial ultraviolet (UV) fixed point/pseudo fixed
point) at high energies, then we have a {\em power
enhancement} instead of the logarithmic one in QCD:
\begin{equation}
 \exp \left[ \int_0^t \gamma_m(t') dt' \right]
    \simeq e^{\gamma_m t}
    = \left( \frac{-p^2}{\mu^2}
      \right)^{\gamma_m/2}.
\label{eq:(4)}
\end{equation}
Accordingly, we have power-enhanced order parameters
for the generic fermion $\psi$:
\begin{eqnarray}
 \Sigma(p^2) &\simeq& \frac{\VEV{\bar \psi \psi}_\mu}{p^2}
            \cdot \left( \frac{-p^2}{\mu^2} \right)^{\gamma_m/2},
\label{power-damp} \\
 \VEV{\bar \psi \psi}_\Lambda 
  &=&  Z_m^{-1}\VEV{\bar \psi \psi}_{\mu}
  \simeq\left( \frac{\Lambda}{\mu}
              \right)^{\gamma_m}
            \cdot    \VEV{\bar \psi \psi}_{\mu}.
\label{eq:(5)}
\end{eqnarray}  
This is actually the mechanism that Holdom\cite{kn:Hold81} proposed 
{\em without explicit dynamics} to resolve the problems of FCNC 
and the light pseudo
NG bosons in technicolor, by simply assuming $\gamma_m\geq 1$.
  
\section{Technicolor}

\subsection{Scaling-Up QCD}

Technicolor (TC)\cite{kn:FS81}
 is replacing the elementary Higgs fields in the SM by 
composite mesons, analogues of $\pi$ and $\sigma$ in QCD, which
are made out of hypothetical fermions called technifermions interacting
via hypothetical gauge interactions with the gauge bosons called 
technigluons. Actually,
the Higgs Lagrangian in SM
\begin{equation}
{\cal L}_{\rm Higgs}= |\partial_{\mu}\phi|^2-\lambda_4
\left(|\phi|^2-\frac{1}{2}v^2\right)^2
\label{Higgs}
\end{equation}
 is precisely the same as the bosonic
part of the $\sigma$ model in (\ref{sigmamodel}), when we rewrite
$\phi$ as
\begin{equation}
 \phi =\frac{1}{\sqrt{2}}
  \left(
\begin{array}{c}
i\pi_1+\pi_2\\
\sigma-i\pi_3
\end{array}
\right) .
\end{equation}
The transformation property can easily be checked through the identification
of $2\times 2$ matrix $M$ in (\ref{Mdef}) with $M=(\tilde \phi, \phi)$,
where $\tilde \phi=i \tau_2 \phi^*$: $\phi \rightarrow g_L\,\phi$,
$\tilde\phi \rightarrow g_L\, \tilde\phi$. The only difference is the 
scale of the order parameters: $v=F_{\pi}=250 {\rm GeV}$
in the Higgs Lagrangian in contrast to $v=f_{\pi}=93 {\rm MeV}$ in the $\sigma$ 
model for hadrons, roughly 2600 times larger.

Then the simplest idea to regard the Higgs fields $\phi$ as composites 
would be to scale-up QCD: $\Lambda_{QCD} \rightarrow
\Lambda_{TC} \simeq 2600 \Lambda_{QCD}$. Thus the low energy limit
of the TC is precisely described by the Higgs Lagrangian (\ref{Higgs}).
When we switch on the $SU(2)_L
\times U(1)_Y$ gauge interactions to the technifermion doublet $\psi$,
the composite NG bosons $\pi^a\sim \bar i\psi\gamma_5 \tau^a \psi$ are absorbed into $W$ and $Z$ bosons through dynamical Higgs mechanism\cite{kn:FS81}:
\begin{equation}
  m_W^2=\left({g_2 \over 2}F_{\pi^{\pm}}\right)^2,
  \qquad       m_Z^2 \cos^2\theta_W =\left({g_2 \over 2}F_{\pi^0}\right)^2,
\label{eq:W/Zmass}
\end{equation}
  where $g_2$ is the $SU(2)_L$ gauge coupling. $F_{\pi^{\pm}},
  F_{\pi^0} \simeq 250{\rm GeV}$ are the decay constants
    of the composite NG bosons $\pi^{\pm},\pi^{0}$ to be absorbed
     into $W$ and $Z$ bosons, respectively, and determine the scale of
technifermion dynamical mass, or the technifermion condensate.  
 
\subsection{Need for Large Anomalous Dimension}
 
 This is a beautiful idea to account for the origin of $W, Z$ boson
 masses.
 What about the mass of quarks and leptons, then?
 We would need some interactions to communicate the composite Higgs sector
 (technifermion condensate) to the quarks and leptons, analogues of the Yukawa 
 interactions in $\sigma$ model 
 and SM itself. Extended TC (ETC)\cite{kn:FS81} 
would be the simplest idea to give
 rise to such (effective four-fermion) interactions among technifermions and
 quarks/leptons through the ETC gauge symmetry. ETC unifies the 
 technifermions and quarks/leptons into the same multiplets and then
 split them in the course of spontaneous symmetry breaking of the ETC gauge
 symmetry at somewhat higher scales called ETC scales $\Lambda_{ETC}
 \gg \Lambda_{TC}$. The effective four-fermion couplings between
 the quark/lepton mass operators and the technifermion condensates  
$\VEV{\bar \psi \psi}$
 are characterized by the ETC scales as $1/\Lambda_{ETC}^2$.
 Then the
quarks/leptons masses are 
given by
\begin{equation}
m \simeq \frac{-\VEV{\bar \psi \psi}_{\Lambda_{ETC}}}{{\Lambda_{ETC}}^2}.
\label{ETCmass}
\end{equation}

If we were assuming QCD-like theories, then (\ref{QCDcond-ren}) would imply
$\VEV{\bar \psi \psi}_{\Lambda_{ETC}} \simeq
\VEV{\bar \psi \psi}_{\Lambda_{TC}}$ up to small logarithmic
corrections. This is disastrous, since ETC unification also produces 
FCNC with strength of the same order, e.g.,  $\sim 
({\bar s}_L\gamma_{\mu} d_L)^2/\Lambda_{ETC}^2$,
which is, however, bounded by 
the experiments of $K^0-\bar K^0$ mixing   
as $1/\Lambda_{ETC}^2< 10^{-5} {\rm TeV}^{-2}$ in 
the case of $s$ quark. Since $\VEV{\bar \psi \psi}_{\Lambda_{TC}}$
$\simeq -(\Lambda_{TC})^3 \simeq -(250 {\rm GeV})^3$,
or $\Lambda_{TC}/\Lambda_{ETC}<10^{-3}$ in the typical TC model,
(\ref{ETCmass}) implies at most $m_s \simeq 0.1 {\rm MeV}$, 
i.e., $10^{-3}$ smaller than the 
reality. Thus the TC as a naive QCD scale-up was dead in the early 
80's.\cite{kn:FS81}

However, the situation is drastically changed in 
the theory with large anomalous dimension
 $\gamma_m \geq 1.$\cite{kn:Hold81} 
 Through(\ref{eq:(5)}) such a theory will actually yield a large
 enhancement factor $10^3$ as desired:
\begin{equation}
 \frac{\VEV{\bar \psi \psi}_{\Lambda_{ETC}}}
{\VEV{\bar \psi \psi}_{\Lambda_{TC}}} 
  \simeq \left( \frac{\Lambda_{ETC}}{\Lambda_{TC}}
              \right)^{\gamma_m}
            \geq \frac{\Lambda_{ETC}}{\Lambda_{TC}}
            >10^3 .
\end{equation}

There is another 
syndrome of TC (not ETC), namely,  the typical TC models 
(``one-family model'') predict many pseudo NG bosons (technipions) in
several GeV region, which are already ruled out by experiments.\cite{kn:FS81}
However, the same enhancement factor due to anomalous
dimension simultaneously resolves this problem by
raising their masses to $O(\Lambda_{TC})$:\cite{kn:Hold81}
\begin{equation}
{m_{\rm pNG}}^2 \sim  
\frac{ (\VEV{\bar \psi \psi}_{\Lambda_{ETC}})^2
      }
{F_{\pi}^2\Lambda_{ETC}^2}
\geq\frac{(\VEV{\bar \psi \psi}_{\Lambda_{TC}})^2}
{\Lambda_{TC}^4}\simeq \Lambda_{TC}^2 ,
\end{equation}
where $F_{\pi}\simeq \Lambda_{TC}$ was granted.
Thus the large anomalous dimension amplifies a small symmetry violation
(explicit chiral symmetry breaking, etc.) to the full strength through
the enhancement factor in the condensate.
 
\subsection{Other Aspects of Large Anomalous Dimension}
 
Such a large enhancement also amplifies the small symmetry violation of high
energy parameters not only in the condensate but in the mass function itself
particularly for the case $\gamma_m >1$ ($\not \simeq 1$).
This fact was first utilized by 
MTY\cite{kn:MTY89a,kn:MTY89b,kn:Yama92} based on an explicit
dynamics in the proposal of a top quark condensate
($m_t \gg m_{b,c,\cdot\cdot\cdot}$), 
and was later re-emphasized in a slightly different 
context\cite{kn:NNT91,kn:BKS91}.

Large anomalous dimension actually implies a tightly bound
NG bosons due to relatively short distance dynamics.
In fact, from (\ref{WT-QMass}) and (\ref{power-damp}) 
we have the high energy behavior 
of the amputated BS amplitude
of the NG bosons at zero NG-boson-momentum:
\begin{equation}
 \chi_\pi^a(p,0) = \frac{1}{F_\pi}\gamma_5 \tau^a\Sigma(-p^2)
 \sim \left( \frac{-p^2}{\mu^2} \right)^{-1+\frac{1}{2}\gamma_m}.
\end{equation}
In QCD with $\gamma_m
\simeq 0$ we find $\chi_\pi \sim (-p^2/\mu^2)^{-1}$ 
and hence the radius of the interaction within the 
composite $\VEV{r} \simeq \mu^{-1} \simeq
F_\pi^{-1}$, in which case the $\sigma$ model description 
obviously breaks down at the order
of $O(F_\pi)$. On the other hand, in the extreme case of $\gamma_m \simeq 2$
we have $\chi_\pi \sim {\rm const.}$ and hence $\VEV{r} \simeq \Lambda^{-1}$
(almost point-like interaction range, or very tightly bound), 
in which case the $\sigma$ model
description persists up to the high energy scale $\Lambda$.

\section{Nambu-Jona-Lasinio Model}

Do such explicit dynamical models as have large anomalous dimension really
exist? The answer is yes. We shall explain solutions of ladder SD equation
for the gauged NJL models (gauge theories plus four-fermion theories)
which actually have large anomalous dimension $1<\gamma_m <2$. 
They encompass a variety of tightly bound composite Higgs models, 
such as walking TC ($\gamma_m \simeq 1$), strong-ETC TC ($1<\gamma_m <2$)
and top quark condensate ($\gamma_m \simeq 2$), etc..

Here we start with discussion on the NJL model\cite{kn:NJL61}.
The  NJL model is of course 
non-renormalizable and trivial theory, i.e., we cannot take the UV cutoff
to infinity to have a sensible continuum theory, in contrast to
the gauged NJL model. Nevertheless it has 
important common features with the gauged NJL model and is a pedagogical
tool to understand physics of the large anomalous dimension.

\subsection{Gap Equation}

Let us consider an NJL model invariant under chiral $SU(2)_L \times SU(2)_R$
transformation:
\begin{equation}
{\cal L}_{\rm NJL}=
\bar \psi i\xbar{\partial}\psi 
+\frac{1}{2N}G 
\left[ \left(\bar \psi \psi\right)^2
+ \left(\bar \psi i \gamma_5 \tau^a\psi\right)^2 
\right],
\label{NJL}
\end{equation}
where $G$ is a four-fermion coupling and $\psi$ is an $N$-component 
fermion. In the large $N$ limit, we have a
famous gap equation corresponding to the Hartree-Fock self-consistent 
equation for the fermion dynamical mass
 $\Sigma (-p^2)=m^*={\rm const.}$:
$ (1/2N) G \left(\bar \psi \psi\right)^2$ $\Rightarrow (G/N) 
\VEV{\bar \psi \psi} \bar \psi \psi=-m^*\bar \psi \psi$. 
(Hereafter we will use $m$ instead of $m^*$ for the
dynamical mass; $m^*\Rightarrow m$.) This yields a famous 
gap equation\cite{kn:NJL61}:
\begin{eqnarray}
m
&=& -\frac{G}{N}\VEV{\bar \psi \psi}=\frac{G}{N}Tr S(p)
    = 4G\int \frac{d^4 p}{(2\pi)^4 i}\frac{m}{m^2-p^2}  \nonumber \\ 
&=& m\cdot \frac{G}{4\pi^2} 
   \left(\Lambda^2 - m^2 \ln \frac{\Lambda^2}{m^2}\right),
\label{gapNJL}
\end{eqnarray}
where we introduced a UV cutoff $\Lambda$ for the Euclidean momentum 
$p^2<\Lambda^2$ (Hereafter we will use Euclidean momentum $-p^2 
\Rightarrow p^2 (>0)$).

There are two solutions to (\ref{gapNJL}):\\
(i) $m=0$ (unbroken solution), \\
(ii) $m\ne0$ such that $SU(2)_L \times SU(2)_R\rightarrow SU(2)_V$
(\SxSB solution), in which case (\ref{gapNJL}) can be rewritten as
\begin{equation}
\ln\frac{\Lambda^2}{m^2}\cdot
\left(\frac{m}{\Lambda}\right)^2 \simeq \frac{1}{g^*}- \frac{1}{g}\: ,
\label{scaling-NJL0}
\end{equation}
where $g\equiv G\Lambda^2/(4\pi^2)$ and $g^*=1$.
We shall refer to this relation as a scaling relation hereafter.  
While the first solution always exist
for any coupling, the second one does only for strong coupling $g\equiv 
\frac{GN\Lambda^2}{4\pi^2} >g^*=1$ for which
 the second solution is more stable 
than the first one. Thus the system is in Wigner phase (i) at 
$g<g^*$ while in NG phase (ii) at $g>g^*$. Both of the phases
 are connected to each other continuously at the critical point $g=g^*=1$: 
Namely, a second order phase transition takes place.
Near the vicinity of the critical coupling $g^*=1$, we may ignore the 
logarithmic factor in (\ref{scaling-NJL0}):
\begin{equation}
\left(\frac{m}{\Lambda}\right)^2 \simeq \frac{1}{g^*}- \frac{1}{g}\: .
\label{scaling-NJL}
\end{equation}

First we note   
that the mass rises sharply from zero to $O(\Lambda)$
as we move the coupling from $g<g^*=1$ to $g>g^*$.
This implies that if different fermions have different $g$'s,  with $g>g^*$ for some and  $g<g^*$ for others, although on the same order of $O(1)$, 
then the former fermions acquire large mass while the latter ones remain 
massless. 
Thus the small asymmetry among couplings 
results in large difference in fermion masses.
This {\it amplification of symmetry violation}
is a salient feature of the critical phenomenon and was first used by
MTY\cite{kn:MTY89a,kn:MTY89b} in the 
proposal of the top quark condensate. 

There are two ways to look at the scaling relation (\ref{scaling-NJL})
in approaching the critical point $g=g^*$: Since
the dimensionless coupling $g$ is a function of only
the ratio $\Lambda/m$, we may regard $\Lambda/m \rightarrow \infty$ as either
the limit of $m \rightarrow 0$ with $\Lambda ={\rm fixed}$ or
that of $\Lambda \rightarrow \infty$ with $m={\rm fixed}$.
In the first picture we stay in the 
cutoff theory and requires {\it fine-tuning} of the coupling $1/g^*-1/g
\ll 1$ in order to realize the hierarchy $m\ll \Lambda$.

\subsection{Continuum Limit, or Renormalization}

In the second picture, on the other hand, we take the continuum limit
$\Lambda \rightarrow \infty$ according to the RGE point of view. The couling is now required to get renormalized  and runs depending on
the cutoff $\Lambda$, with the beta function being calculated from 
 (\ref{scaling-NJL}): 
\begin{equation}
\beta (g) \equiv 
\Lambda \frac{\partial}{\partial \Lambda} g(\Lambda)
 = 2 g \left(1 - \frac{g}{g^{*}} \right).
\label{eq:betaNJL} 
\end{equation}
This implies existence of a nontrivial UV fixed point at the critical point:
 $g(\Lambda) \rightarrow g^*$ as $\Lambda \rightarrow \infty$. 
From (\ref{gapNJL}) the condensate reads:
\begin{equation}
\VEV{\bar \psi \psi}_{\Lambda}
=-Tr S(p) \simeq -\frac{mN}{4\pi^2} \Lambda^2
\simeq \left(\frac{\Lambda}{m}\right)^2 \cdot \VEV{\bar \psi \psi}_{m}
\equiv Z_m^{-1}\VEV{\bar \psi \psi}_{m}
\label{cond-NJL}
\end{equation}
up to logarithm.
Then we have a large anomalous dimension at the UV fixed point:\cite{kn:KY90}
\begin{equation}
 \gamma_m (g=g^*)
 \equiv -\Lambda \frac{\partial}{\partial \Lambda} \ln Z_m
 = 2.
\label{eq:gammaNJL}
\end{equation}
Thus the ``fine-tuning'' has been traded for the RGE concepts, namely, 
a nontrivial UV fixed point 
and large anomalous dimension. 

It should be mentioned that 
the above RGE arguments are only {\it formal}, since the NJL 
model is well known to be a {\it nonrenormalizable}
and {\it trivial} theory. The above renormalization procedure only removed
the quadratic divergence, while the {\it logarithmic} one in 
(\ref{scaling-NJL0}) is actually a trouble for the renormalization.
As we explicitly do in the later section, we may write the NJL model into 
the form of the Yukawa
 model, consisting of the original fermion and the auxiliary scalar field 
$\phi$ which have no kinetic term $(\partial_{\mu}\phi)^2$ and no quartic 
coupling $\lambda_4 \phi^4$. These terms are actually induced by the fermion 
loop  through the Yukawa 
coupling and are logarithmically divergent 
$\sim \ln \Lambda$ as in the case of the ordinary Yukawa model.
In contrast to the Yukawa model, however, 
the NJL model has {\it no counter terms} like 
$(\bar\psi\psi)^4$ (quartic coupling of $\phi$) and 
$
  \partial_\mu (\bar\psi\psi) \partial^\mu (\bar\psi\psi)
$ (kinetic term of $\phi$)  in the
original Lagrangian and hence 
these logarithmic divergences cannot be renormalized.
Thus the NJL model is {\it not renormalizable}.
We may rescale the induced kinetic term of $\phi$ into the usual one by
rescaling $\phi$, then the logarithmic divergence moves over to
the rescaled Yukawa coupling (effective Yukawa coupling) which now behaves as 
$(1/\ln\Lambda)^{1/2}\rightarrow 0$ as $\Lambda \rightarrow \infty$.
This can also be seen from the PS formula for $f_\pi^2$
(exact at $1/N$ leading order in the NJL model),
(\ref{PS-QCD}), which is
logarithmically divergent since $\Sigma (p^2) \equiv m$. Accordingly,
the effective Yukawa coupling $y\equiv \sqrt{2} m/f_\pi$ vanishes as the above.
This is nothing but a {\it triviality} of the NJL model. Then the above 
UV fixed point is a Gaussian fixed point (free theory) and does not produces
a sensible interacting continuum theory.

\subsection{Renormalization in $D(2<D<4)$ Dimensions}

Nevertheless, it was shown\cite{kn:KY90,kn:HKWY92,kn:KTY93} that 
the above characteristic features of the NJL model,
 the UV fixed point and large anomalous 
dimension, become true for the NJL model in $D (2<D<4)$ dimensions 
(${\rm NJL}_{<4}$) 
which are known to be 
{\it renormalizable} and {\it nontrivial} in the continuum limit 
$\Lambda \rightarrow \infty$.\cite{kn:RWP90} 
The fine-tuning of ${\rm NJL}_{<4}$ model 
is in fact connected with a sensible
continuum theory, in sharp contrast to the NJL model.
 
Actually, in $D$ dimensions 
the scaling relation (\ref{scaling-NJL0}) and the condensate (\ref{cond-NJL})
become\cite{kn:KTY93}
\begin{eqnarray}
\frac{2}{4-D}\cdot
\left(\frac{m}{\Lambda}\right)^{D-2} 
&\simeq& \frac{1}{g^*}- \frac{1}{g}\: ,
\label{scaling-NJLD} \\
\VEV{\bar \psi \psi}_{\Lambda}
&\simeq& \left(\frac{\Lambda}{m}\right)^{D-2} \cdot \VEV{\bar \psi \psi}_m
\label{cond-NJLD}
\end{eqnarray}
at $1/N$ leading order,
where $g^*=D/2-1$. Similarly to (\ref{eq:betaNJL}) and (\ref{eq:gammaNJL}),
RGE functions can be obtained from (\ref{scaling-NJLD}) and
(\ref{cond-NJLD}): 
\begin{eqnarray}
 \beta(g)   &=& (D-2) g
\left(1 - \dfrac{g}{g^{*}} \right),\nonumber\\
\gamma_m 
&=& D-2.
\label{RGEfnD}
\end{eqnarray}
The factor $1/(2-D/2)$ in L.H.S. of  (\ref{scaling-NJLD}) 
reflects the logarithmic divergence of (\ref{scaling-NJL0})
in the limit $D\rightarrow 4$.
This  {\it absence of the logarithmic divergence} in $D (2<D<4)$ dimensions
in contrast to $D=4$ is also true for
the induced kinetic term and
the induced 
quartic coupling of $\phi$ in ${\rm NJL}_{<4}$ model, which are indeed
{\it finite} at $\Lambda \rightarrow \infty$ due to the $D (<4)$ dimensional 
momentum integral.

Then we can perform explicit renormalization
of the NJL$_{<4}$ model by simply removing power divergences
in the effective 
potential.\cite{kn:HKWY92,kn:KTY93}
This leads to the RGE functions,
$\beta (g)$ and $\gamma _m (g)$, written in terms of the {\it renormalized 
coupling} $g$ in the {\it continuum} theory 
($\Lambda \rightarrow \infty$):\cite{kn:KY90,kn:HKWY92,kn:KTY93}
\begin{eqnarray}
  \beta(g)   &=& (D-2) g
\left(1 - \dfrac{g}{g^{*}} \right),
\label{eq:(5.10)}  \nonumber \\
  \gamma _m(g) &=& (D-2) \dfrac{g}{g^{*}}\:,
\label{eq:(5.11)}
\end{eqnarray}
where $\beta \equiv(\mu \partial /\partial\mu) g$ and 
$\gamma _m  \equiv (\mu \partial/\partial \mu) \ln Z_{\mu} $,
with $\mu $ being the renormalization point.
The above expressions are obtained from 
effective potential and hence valid both in the
unbroken and \SxSB phases. 
These RGE functions take the same form in the bare coupling, which coincide
with (\ref{RGEfnD}) obtained from the gap equation (scaling relation)
and the condensate at $g\simeq g^*$ in the \SxSB phase. 

In terms of the anomalous dimension, we may regard the renormalizability of the NJL$_{<4}$ model as follows:
The fact that we can renormalize the theory without higher dimensional
operators $(\bar\psi\psi)^4$ (quartic coupling of $\phi$) and 
\(
  \partial_\mu (\bar\psi\psi) \partial^\mu (\bar\psi\psi)
\) (kinetic term of $\phi$) 
at $1/N$ leading order simply reflects   the following fact:
$(\bar \psi \psi)^2$ is a {\it relevant operator} due to a large anomalous 
dimension $\gamma_m=D-2$ at $g=g^{*}$, i.e., 
$\dim(\bar \psi \psi)^2=2(D-1-\gamma_m)=2<D$,
while the would-be ``counter terms'' $(\bar \psi \psi)^4$ and
\(
  \partial_\mu (\bar \psi \psi) \partial^\mu (\bar \psi \psi)
\) are {\it irrelevant operators}, 
\(
  \dim(\bar \psi \psi)^4=4(D-1-\gamma_m)=4>D
\),
\(
 \dim[\partial_\mu (\bar \psi \psi) \partial^\mu (\bar \psi \psi)]
  =2(D-\gamma_m)=4>D
\).
At $D=4$, however, all these operators equally have dimension $4(=D)$ 
and become {\it marginal operators}.  Hence they should be included in order
to make the theory renormalizable, in which case the NJL model in its
renormalizable version becomes identical to the Higgs-Yukawa 
 system
($\sigma$ model, or "standard model")\cite{kn:Suzu90}.
\par

\section{Gauged Nambu-Jona-Lasinio Model}

Now we discuss  the gauged NJL model in four dimensions which 
was shown to be renormalizable due to large anomalous 
dimension, $1<\gamma_m<2$, in a sense similar  to the
renormalizability of 
NJL$_{<4}$.\cite{kn:KSY91,kn:KTY93,kn:Yama92,kn:Kras93,kn:HKKN94}

\subsection{Ladder Schwinger-Dyson Equation and Critical Line}

Let us start with the Lagrangian of the gauged NJL model, 
the NJL model (\ref{NJL}) plus $SU(N)$ gauge theory:
\begin{equation}
  {\cal L} = 
    \bar\psi  (i \fsl{\partial}-e \fsl{A}) \psi 
   +\frac{G}{2N} \left[ 
                  (\bar\psi \psi )^2 + (\bar\psi i\gamma _5 \tau^a\psi)^2
                \right]
   -\frac{1}{2} Tr(F_{\mu \nu }F^{\mu \nu }),
\label{GNJL}
\end{equation}
where
$e$ is the gauge coupling constant. Here we first discuss the
case of non-running gauge coupling (``standing'' limit of 
walking gauge coupling).
In the ladder approximation the SD equation takes the same form as
that of the QED plus NJL model.
In Euclidean space, the ladder SD equation for the fermion   
propagator $iS^{-1}(p)=Z_{\psi}^{-1}(p^2)(\xbar{p} -\Sigma(p^2)) $
in Landau gauge takes the form (after angular integration):
\begin{equation}
  \Sigma(x)
  = {g \over \Lambda^2} \int_0^{\Lambda^2} dy 
      {y\Sigma(y) \over y+\Sigma(y)^2}
    +\int_0^{\Lambda^2} dy
      {y\Sigma(y) \over y+\Sigma(y)^2} K(x,y),
\label{eq:(4.2)}
\end{equation}
where $x \equiv p^2 $, $K(x,y) =\lambda/\max(x,y)$ and
$
\lambda \equiv (3C_2(\mbox{\boldmath$F$})/4\pi)(e^2/(4\pi))$,
with 
$
C_2(\mbox{\boldmath$F$})=(N^2-1)/2N
$
being  
the quadratic Casimir of the fermion representation 
$\mbox{\boldmath$F$} (= \mbox{\boldmath$N$})$, respectively.
(Note that $Z_{\psi}(p^2)\equiv 1$ in Landau gauge in the ladder approximation.)
The dynamical mass function is normalized as $\Sigma(m^2) = m$ as before.

Eq.(\ref{eq:(4.2)})
 was first studied by Bardeen, Leung and
Love\cite{kn:BLL86} in QED for the strong gauge coupling region
$\lambda>\lambda_c =1/4$. A full set of \SxSB solutions 
in the whole ($\lambda,g$) plane and the {\it critical line}
were found by Kondo, Mino and Yamawaki and
independently by Appelquist, Soldate, Takeuchi and
Wijewardhana.\cite{kn:KMY89} 
In particular, at $0<\lambda<\lambda_c=1/4$ 
the asymptotic form of the solution of the ladder SD 
equation (\ref{eq:(4.2)}) takes the form:\cite{kn:KMY89}
\begin{equation}
  \Sigma(p^2)\mathop{\simeq}_{p\gg m} 
    \simeq m\left({p\over m}\right)^{-1+\omega}
 \qquad (0< \lambda < \lambda_c), 
\label{eq:(4.3)}
\end{equation} 
which is reduced to a constant mass function $\Sigma (p^2)\equiv 
{\rm const.}=m$ in the pure NJL limit ($\lambda \rightarrow 0$) as it
should be. At $\lambda =\lambda_c$  
we have $\Sigma (p^2) \sim 1/p$ $ (\omega =0)$ up to logarithm 
(See Section 6.1). 

The critical line in the ($\lambda,g$) plane is a generalization of the
 critical coupling in NJL model. It is the line of the second-order phase
transition separating spontaneously broken ($m/\Lambda\ne0$)
and unbroken ($m/\Lambda=0$) phases of the chiral symmetry
(Fig.1)\cite{kn:KMY89}:
\par
\begin{figure}
\begin{center}
\critline \\
\begin{minipage}{5in}
  {\footnotesize
    {\bf Fig.1 \quad} 
  Critical line in $(\lambda,g)$ plane. It 
  separates spontaneously broken (\SxSB) phase and
  unbroken phase ($Sym.$) of the chiral symmetry.
}
\end{minipage}
\end{center}
\end{figure}
\begin{eqnarray}
  g
  &=&{1 \over 4}(1+\omega)^2
  \equiv g^{*},
  \qquad
  \omega \equiv \sqrt{1-\frac{\lambda}{\lambda_c}}
  \qquad (0< \lambda < \lambda_c), 
   \nonumber \\
  \lambda
  &=& \lambda_c =\frac{1}{4} \qquad (g<{1\over4}). 
\label{eq:(2.17)}
\end{eqnarray}

Here the overall mass scale $m$
at $0<\lambda<\lambda_c$ satisfies the scaling relation similar to 
(\ref{scaling-NJL}):\cite{kn:KMY89}
\begin{equation}
\frac{2}{1-\omega^2}\cdot
\left({m \over \Lambda}\right)^{2\omega} \simeq \frac{1}{g^*}
 - \frac{1}{g}\:,
\label{scalinggNJL}
\end{equation}
where the factor $2/(1-\omega^2)$ in L.H.S. gives a divergence in the pure
NJL limit $\omega \rightarrow 1 (\lambda \rightarrow 0)$, which actually
corresponds to the logarithmic divergence in the pure NJL model,
(\ref{scaling-NJL0}). (A careful
analysis at $\omega \rightarrow 1$ in fact yields logarithmic 
divergence)\cite{kn:NSY89}.
Again the absence of this logarithmic factor at $\omega\ne 1 (\lambda \ne 0)$
is relevant to the renormalizability of the gauged NJL model. 

As in the pure NJL model Eq.(\ref{scalinggNJL}) indicates
that the dynamical mass $m$  sharply rises
as we move away from the critical coupling $g=g^*$. 
Now the critical coupling  
on the critical line $g=g^*(\lambda)$ does depend on
the value of gauge coupling $\lambda$, and vice versa $\lambda =\lambda^*(g)$. 
This again means that {\it even 
a tiny difference (symmetry violation) of $\lambda$ (or $g$) for the same $g$
(or $\lambda$) can cause amplified effects on the dynamical mass; $m=0$
(below the critical line) or $m\ne 0$ (above the critical line).}

\subsection{Large Anomalous Dimension and Renormalizability/Nontriviality}

Again the scaling relation (\ref{scalinggNJL}) leads to the beta function
\begin{equation}
\beta(g) = 2\omega g \left(1-\frac{g}{g^*}\right)
\qquad (g\simeq g^*),
\label{betabare}
\end{equation}
while the solution (\ref{eq:(4.3)}) yields the condensate
\begin{equation}
\VEV{\bar \psi \psi}_{\Lambda}= -TrS(p)=-\frac{N}{4\pi^2}
\int_0^{\Lambda^2} d x \frac{x \Sigma(x)}
{x+\Sigma(x)^2}
\simeq \left(\frac{\Lambda}{m}\right)^{1+\omega}
 \cdot \VEV{\bar \psi \psi}_{m},
\label{condgNJL}
\end{equation}
 which implies $Z_m^{-1}\sim \Lambda^{1+\omega}$ and hence 
a large anomalous dimension:\cite{kn:MY89}
\begin{equation}
\gamma_m =1+\omega \qquad (0< \omega \equiv 
\sqrt{1-\frac{\lambda}{\lambda_c}}< 1)
\label{eq:gammam}
\end{equation}
at the critical line.
In particular, we reproduce $\gamma_m = 2$ in the pure NJL model ($\lambda =0$)
and $\gamma_m =1$ for $\lambda =\lambda_c$.
The solution of the SD equation (\ref{eq:(4.3)})
is  also compared with the general result 
from OPE and RGE, (\ref{power-damp}):  
\begin{equation}
  \Sigma(p^2)\mathop{\simeq}_{p\gg m} 
   {m^3\over p^2} \left({p\over m}\right)^{\gamma_m},
\end{equation}
where we have set $\mu=m$ and $\VEV{\bar \psi \psi}_{\mu}
 \simeq -m^3$. Then we find that
 such a slowly damping solution (\ref{eq:(4.3)}) 
actually corresponds to 
 a large anomalous dimension (\ref{eq:gammam}).

As in NJL$_{<4}$ model, induced kinetic term and induced quartic 
coupling of the auxiliary field 
$\phi$ in the gauged NJL model 
remain {\it finite} in the continuum limit $\Lambda \rightarrow 
\infty$, this time thanks to the power damping behavior of the mass 
function (\ref{eq:(4.3)})
in contrast to the constant mass function in the pure NJL model.
Such a damping behavior is due to the presence of gauge interactions.  
This finiteness in turn implies the finiteness of the effective Yukawa coupling
in the continuum limit, namely, the {\it nontriviality}
 of the gauged NJL model. This can also be seen from the PS formula 
(\ref{PS-QCD}) whose integral is now convergent thanks to the power-damping 
behavior of $\Sigma(p^2)$.

Actually, an explicit renormalization
procedure in the ladder approximation
was performed by Kondo, Tanabashi and 
Yamawaki\cite{kn:KTY93} through the effective 
potential as in NJL$_{<4}$.
The fine-tuning of the bare couling $1/g^*-1/g\ll 1$
in (\ref{scalinggNJL}) corresponds to the continuum limit 
$\Lambda/m \rightarrow \infty$, which now defines a finite
renormalized theory explicitly written in terms
of renormalized quantities, in sharp contrast to
the pure NJL model where a similar fine-tuning through (\ref{scaling-NJL0})
or (\ref{scaling-NJL}) has nothing to do
with a finite renormalized theory. 
This renormalization leads to the beta function and the anomalous 
dimension:\cite{kn:KTY93}
\begin{eqnarray}
  \beta(g) &=& 2 \omega g \left(1-\frac{g}{g^{*}} \right), 
\label{eq:beta}  \\
  \gamma_m(g) &=& 1-\omega + 2\omega \frac{g}{g^{*}}
\label{eq:gamma}  
\end{eqnarray}
at $0< \lambda < \lambda_c$, 
 where $g$ is {\it either renormalized or bare} coupling. 
These expressions are valid both in the \SxSB and unbroken phases.
Note that these RGE functions at $g \simeq g^*$ take the same form as 
(\ref{betabare}) and (\ref{eq:gammam})
obtained from the scaling relation and the condensate in \SxSB phase.
It is now clear that the critical line $g=g^*=\frac{1}{4}(1+\omega)^2$ is 
a UV fixed line where the anomalous dimension takes the 
large value
\begin{equation}
1<\gamma_m(g=g^*) = 1+\omega<2.
\end{equation}

The essence of the renormalizability now resides in the fact
that this dynamics possesses a large anomalous dimension
$\gamma_m>1$ but not too large, $\gamma_m <2$.\cite{kn:KTY93}
It in fact implies that 
 the four-fermion interactions are {\it relevant operators},
  $2< d_{(\bar \psi \psi)^2}=2(3-\gamma_m)=4-2\omega <4$.\cite{kn:MY89}
Accordingly, possible {\it higher dimensional interactions},
  $(\bar \psi \psi)^4$, $\partial_\mu(\bar \psi  \psi)
  \partial^{\mu}(\bar \psi  \psi)$, etc.,
are {\it irrelevant operators} ($d >4$ due to $d_{\bar \psi \psi}>1$), 
 in contrast to the pure NJL model ($\omega=1$)
where these operators
are marginal ones ($d=4$ due to $d_{\bar \psi \psi}=1$)).
 Thus the {\it presence of the gauge interactions} can change
 drastically the four-fermion theories into {\it renormalizable theories}
 without introducing ``higher dimensional 
operators''.\cite{kn:KSY91,kn:KTY93,kn:Yama92,kn:Kras93,kn:KSTY94,kn:HKKN94}
These higher dimensional operators are in fact calculated to be finite
and hence no counter terms are needed.

\subsection{Running Effects of Gauge Coupling}

One can easily take account of perturbative 
{\it running effects} of the $SU(N)$ gauge 
 coupling, typically the QCD coupling,
 in the ladder SD equation (``improved ladder SD 
equation'')\cite{kn:Higa84}
by simply replacing $\lambda$ in (\ref{eq:(4.2)}) by the one-loop running
 one $\lambda(p^2)$ parameterized as follows:
\begin{equation}
  \lambda(p^2) 
  =\cases{
    \lambda_{\mu} 
       & ($p <\mu_{IR}$) \cr
       \dfrac{A}{4\ln{( p/  \Lambda_{QCD})}}   
       & ($p >\mu_{IR}$) \cr
},
\label{eq:(3.19)}
\end{equation}
where $A =c/b=18C_2(\mbox{\boldmath$F$})/(11N-2N_f)$ ($=24/(33-2N_f)$
for $N=3$)
 and 
$\lambda_{\mu} 
(= (A/4)/\ln{(\mu_{IR}/\Lambda_{QCD})})
$ are constants and $\mu_{IR} (= O(\Lambda_{QCD})$ an
artificial ``IR cutoff'' of otherwise divergent running coupling constant
(We choose $\lambda_{\mu} > 1/4$ so as to trigger the S$\chi$SB already in
the pure QCD). 

Then the SD equation takes the form 
\begin{equation}
  \Sigma(x) = 
     {g \over \Lambda^2} 
      \integ0{\Lambda^2}dy{y \Sigma(y) \over y+\Sigma^2(y)}
    + \integ0{\Lambda^2} dy {y \Sigma(y) \over y+\Sigma^2(y)}
      {\mbox{\boldmath$K$}(x,y)},
\label{eq:(4.4)}
\end{equation}
where ${\mbox{\boldmath$K$}}(x,y) \equiv
 \lambda(\max(x,y,\mu_{IR}^2))/\max(x,y)$. Note that the
 non-running case is regarded as the ``standing'' limit
  $A \rightarrow \infty$ (with $\lambda_{\Lambda} \equiv \lambda(\Lambda^2)$
fixed) of the walking coupling ($A \gg 1$).\cite{kn:BMSY87}
 
The S$\chi$SB solution of (\ref{eq:(4.4)}) is logarithmically
damping\cite{kn:MY89}, 
\begin{equation}
  \Sigma(p^2) \simeq 
  m\cdot \left( \frac{\ln\frac{p}{\Lambda_{QCD}}}
                 {\ln\frac{m}{\Lambda_{QCD}}}
                  \right)^{-\frac{A}{2}},
\label{eq:(4.5)}
\end{equation}
which is essentially the same as 
 (\ref{eq:(4.3)}) with a small power $\lambda (\sim \lambda_{\Lambda})
 \ll 1$. 
In the case of pure QCD ($g=0$), such a very slowly damping solution
 (``irregular asymptotics'') is the {\it explicit} chiral-symmetry-breaking
 solution due to the current quark mass ($\gamma_m \simeq 2\lambda(p^2)
=A/(2{\bar t})\ll 1$,  see (\ref{gammamQCD})).\cite{kn:Lane74,kn:Higa84}
  However, Miransky and
 Yamawaki\cite{kn:MY89}
 pointed out that it can be the {\it \SxSB  solution}
 {\it in the presence of an
 additional four-fermion interaction.} 
Accordingly, the condensate is quadratically enhanced as in pure NJL model
except for a logarithmic correction:
\begin{equation}
\VEV{\bar \psi \psi}_{\Lambda} 
=-\frac{N}{4\pi^2}
\int_0^{\Lambda^2} d x \frac{x \Sigma(x)}
{x+\Sigma(x)^2}
\simeq 
     \left( \frac{\Lambda}{m}\ \right)^2 
     \left( \frac{\ln\frac{\Lambda}{\Lambda_{QCD}}}
                 {\ln\frac{m}{\Lambda_{QCD}}}
                  \right)^{-\frac{A}{2}}
     \cdot \VEV{\bar \psi \psi}_m.
\label{condgNJLrun}
\end{equation}

The solution (\ref{eq:(4.5)}) corresponds to a very
 large anomalous dimension 
\begin{equation}
\gamma_m \simeq 2-2\lambda(p^2),
\label{gammamrun}
\end{equation}
or
\begin{equation}
\exp \left[ \int_0^t \gamma_m(t') dt' \right]
   =\exp \left[2t -\frac{A}{2} \ln (\frac{\bar t}{{\bar t}_\mu})\right] 
= \left( \frac{p}{m}
      \right)^2
\left( \frac{\ln\frac{p}{\Lambda_{QCD}}}
                 {\ln\frac{m}{\Lambda_{QCD}}}
                  \right)^{-\frac{A}{2}},
\end{equation}
(compare with (\ref{eq:(3)})). For $p^2=\Lambda^2$ we have
$\gamma_m (g^*) =2-2 \lambda_{\Lambda}$
near the ``critical line''
 \begin{equation}
 g=g^* \simeq 1-2\lambda_{\Lambda}
\label{eq:``criticalline''}
\end{equation}
at $\lambda_{\Lambda} \ll 1$.\cite{kn:MY89,kn:Take89,kn:KSY91,kn:BKS91}
(There is no
critical line in the rigorous sense in this case, since \SxSB 
takes place in the whole coupling region due to pure QCD dynamics 
which yields dynamical mass $m=m_{QCD}= O(\Lambda_{QCD})$.)  
Note that (\ref{gammamrun}) and 
(\ref{eq:``criticalline''}) coincide with those in the
 non-running case, (\ref{eq:gammam}) and (\ref{eq:(2.17)}), respectively, 
at $\lambda \ll 1$.

As to the renormalizability of the gauged NJL model in this case, 
convergence of the kinetic term and
the quartic self-coupling of $\phi$ is the same as the convergence
of $f_{\pi}$ through the PS formula (\ref{PS-QCD}), which
depend on the power of the logarithmic damping factor in (\ref{eq:(4.5)}),
$\sim \ln p^{-A/2}$. Thus $A>1$ is the condition for the convergence
of $f_\pi$ and hence for the 
renormalizability.%
\cite{kn:KSY91,kn:KTY93,kn:Yama92,kn:Kras93,kn:KSTY94,kn:HKKN94}
This point will be explained further in the Section 7.

\section{Tightly Bound Composite Higgs Models}

Now that we have seen that the gauged NJL model is an explicit dynamics
which has a large anomalous 
dimension and sensible continuum limit, we can discuss possible
applications of it to the electroweak symmetry breaking, namely the 
tightly bound composite 
Higgs models mentioned before. 
There are a variety of tightly bound composite Higgs models
based on the gauged NJL model; walking TC 
($\gamma_m \simeq 1$) \cite{kn:Hold85,kn:YBM86},
strong ETC TC ($1<\gamma_m <2$) \cite{kn:MY89} and 
top quark condensate ($\gamma_m \simeq 2$) 
\cite{kn:MTY89a,kn:MTY89b,kn:Namb89,kn:Marc89,kn:BHL90}, 
etc., whose order parameters are all enhanced 
by the factor
\begin{equation}
\exp \left[ \int_0^t \gamma_m(t') dt' \right],
\end{equation}
where $t$ could be $\ln (p/m)$ or $t_\Lambda(=\ln (\Lambda/m))$ .

\subsection{Walking Technicolor}

It was first pointed out by Yamawaki,
Bando and Matumoto\cite{kn:YBM86} that the technicolor within
the {\em ladder SD equation} (with the gauge coupling
constant {\em fixed}, i.e., non-running) possesses an 
\SxSB\ {\em solution} with
a large anomalous dimension: 
\begin{eqnarray}
 \gamma_m &\simeq& 1,\\
 \Sigma(p^2) &\sim& \frac{1}{p} \qquad (p \gg \Lambda_{TC}),\\
 \VEV{\bar \psi \psi}_{\Lambda_{ETC}} &\simeq&
  \left( \frac{\Lambda_{ETC}}{\Lambda_{TC}} \right)
 \cdot \VEV{\bar \psi \psi}_{\Lambda_{TC}},
\end{eqnarray} 
and hence resolves the long standing problems of the old TC, a scale-up
of QCD, in a sense  discussed in Section 3.
Essentially the same observation was also made    
by Akiba and Yanagida\cite{kn:YBM86} without notion of anomalous
dimension.
It should also be mentioned
that Holdom\cite{kn:Hold85} earlier recognized the same 
dynamics through a purely {\em numerical} analysis of the 
ladder-type SD equation (without notion of anomalous dimension).

The above feature is actually the essence of the ``walking TC'',
a generic name currently used (see Appelquist et al.\cite{kn:YBM86}) 
for a wider class of TC's with $\gamma_m \simeq 1$ due to slowly
running (``walking'', $A \gg 1$) 
gauge coupling {\em including} the non-running 
($A \rightarrow \infty$, ``standing'') case
as an extreme case. In order for the walking TC to be a realistic solution of 
the FCNC problem, however, it must be very close to the standing limit 
anyway.\cite{kn:BMSY87} In the standing limit
the \SxSB\ solution exists only when the gauge coupling $\lambda$ 
exceeds a critical value
$\lambda_c =1/4$. Hence the critical coupling plays a role of a nontrivial
UV fixed point. Thus the walking TC may be viewed as the TC with a
nontrivial UV fixed point/pseudo fixed point, with the coupling being 
kept close to the critical coupling all the way up to $\Lambda_{ETC}$ scale.

Moreover, as we mentioned before,
 $\gamma_m =1$ is realized in the gauged NJL model 
at $\lambda =\lambda_c$. 
Actually, it was suggested\cite{kn:KSY91}
that the non-zero four-fermion coupling $g\simeq 1/4$ at $\lambda =\lambda_c$
might be ``{\it induced}'' by the dynamics of the (standing) gauge theory 
itself. If it is the case, the
standing/walking TC might be realized at $(\lambda,g) =(\lambda_c,1/4)$ but
not at $(\lambda_c,0)$ as was considered originally.

\subsection{Strong ETC Technicolor}

Next we come to the TC with even larger anomalous dimension,
$1 < \gamma_m <2$, which is due to strong four-fermion coupling
$g \simeq g^*$ arising from ETC interaction, Pati-Salam interaction,
preonic interaction, etc.. This is generically dubbed a strong 
ETC technicolor\cite{kn:MY89}.
Based on the \SxSB\ solution\cite{kn:KMY89},
we have an even  bigger enhancement of the order
parameters due to such a large anomalous dimension:
\begin{eqnarray}
 1 < \gamma_m &=& 1+\sqrt{1-\frac{\lambda}{\lambda_c}}<2,\\
 \Sigma(p^2) &\sim& p^{-1+\sqrt{1-\frac{\lambda}{\lambda_c}}},\\
 \VEV{\bar \psi \psi}_{\Lambda_{ETC}} &\simeq& 
     \left( \frac{\Lambda_{ETC}}{\Lambda_{TC}} 
     \right)^{1+ \sqrt{1-\frac{\lambda}{\lambda_c}}}
     \cdot \VEV{\bar \psi \psi}_{\Lambda_{TC}}.
\end{eqnarray}
This {\em in principle} can yield  enhancement of the quark mass
(\ref{ETCmass}),
say, up to $O(\Lambda_{TC})$ and could account for a large
top quark mass (if one takes $\gamma_m \simeq 2$).

\subsection{Towards the Top Quark Condensate}

Once we have taken a TC with such an extremely tightly bound composite Higgs
with $\gamma_m \simeq 2$ in order to accommodate a large top quark 
mass, we may consider a much simpler alternative:
Namely, the top quark itself may play the role of the technifermion
triggering the electroweak symmetry breaking.
In fact, a top quark condensate  was proposed by 
MTY\cite{kn:MTY89a,kn:MTY89b},
based on the \SxSB\ {\em solution of the SD equation} for the
gauged NJL model (this time QCD plus four-fermion interactions) 
with $\lambda_{QCD} \ll 1$ (see (\ref{eq:(4.5)})-(\ref{gammamrun})):  
\begin{eqnarray}
 \gamma_m &\simeq& 1+\sqrt{1-\frac{\lambda_{QCD}}{\lambda_c}}
          \simeq 2- \frac{\lambda_{QCD}}{2\lambda_c}
          \simeq 2- \frac{A}{2 \bar t},\\
 \Sigma(p^2) &\sim& 
     \left( \ln\frac{p}{\Lambda_{QCD}}
                  \right)^{-\frac{A}{2}},\\
 \VEV{\bar t t}_{\Lambda} &\simeq& 
     \left( \frac{\Lambda}{m_t}\ \right)^2 
     \left( \frac{\ln\frac{\Lambda}{\Lambda_{QCD}}}
                 {\ln\frac{m_t}{\Lambda_{QCD}}}
                  \right)^{-\frac{A}{2}}
     \cdot \VEV{\bar t t}_{m_t},
\end{eqnarray}
where $\mu$ was taken as the top quark mass $m_t$.
This model will be explained in somewhat details in the next section.
 
\section{Top Quark Condensate}

\subsection{Why Top Quark Condensate?}

Recently the elusive top quark has been finally discovered and found to have
a mass of about 180 GeV,\cite{kn:TEVATRON} roughly on the order of weak scale 
250 GeV. This is extremely large compared with mass of
all other quarks and leptons and seems to suggest a special role of the top
quark in the electroweak symmetry breaking, {\it the origin of mass},
and hence a strong connection with the Higgs boson itself.  
 
Such a situation can be most naturally understood by the top quark
condensate 
proposed by Miransky, Tanabashi and Yamawaki (MTY)\cite{kn:MTY89a,kn:MTY89b}
 and
by Nambu\cite{kn:Namb89} independently. This entirely replaces
the standard Higgs doublet by a composite one formed by a strongly
coupled short range dynamics (four-fermion interaction) which triggers
the top quark condensate.  The Higgs boson emerges as
a $\bar t t$ bound state and hence is deeply connected with the top
quark itself. Thus the model may be called ``top mode standard
 model''\cite{kn:MTY89b}
  in contrast to the SM (may be called ``Higgs mode standard
 model''). The model was further developed by the renormalization-group 
 (RG) method.\cite{kn:Marc89,kn:BHL90}
 
 Once we understand that the top quark mass is of the weak scale order,
 then the question is why other quarks
 and leptons have very small mass compared with the weak scale. Actually,
 the Yukawa coupling is dimensionless and hence naturally 
 expected to be of $O(1)$. This is the question that MTY 
 \cite{kn:MTY89a,kn:MTY89b} had solved 
  in the top quark condensate through the amplification of the
  symmetry violation in the critical phenomenon.

MTY\cite{kn:MTY89a} introduced explicit 
four-fermion interactions responsible
for the top quark condensate in addition to the standard 
gauge couplings. 
Based on the explicit solution of the ladder
SD equation\cite{kn:KMY89,kn:MY89},
MTY found that even if all the dimensionless four-fermion
couplings are of $O(1)$, only the coupling larger than the critical 
coupling yields non-zero (large) mass, while others do just zero masses.
  This is a salient feature of the {\it critical phenomenon}
or the {\it dynamics with large anomalous dimension}
as was already explained in NJL and gauged NJL models.
Combined with the PS formula\cite{kn:PS79},
MTY predicted
 the top quark mass as a {\it decreasing} function of the cutoff $\Lambda$ and
in particular the minimum value to be about 250 GeV for the Planck
scale cutoff, which actually coincides with the weak 
scale.\footnote{
It should be emphasized that the MTY prediction 
(receipt date: Jan. 3, 1989)\cite{kn:MTY89a} 
was made when the lower bound of the top quark mass through direct experiment
was only 28 GeV (TRISTAN value) and many theorists (including SUSY enthusiasts) were still expecting the value below 100 GeV. 
It in fact appeared absurd at that time
to claim a top mass on the order of weak scale. 
Thus such a large top mass was really a {\it prediction} of the model.
}

The model was further formulated in an elegant
fashion by Bardeen, Hill and Lindner (BHL)\cite{kn:BHL90} in the
SM language, based on the RG equation 
and the compositeness condition. BHL 
essentially incorporates $1/N_c$ sub-leading effects such as
those of the composite Higgs loops and $SU(2)_L\times U(1)_Y$ gauge boson 
loops which were disregarded by the MTY formulation. We can
 explicitly see that BHL is in fact  
{\it equivalent} to MTY at $1/N_c$-leading order. 
Such effects turned out to reduce the above MTY 
value 250 GeV
down to 220 GeV, a somewhat smaller value but still on the
order of the weak scale.  

Although the prediction appears to be
substantially higher than the experimental value  mentioned
above\cite{kn:TEVATRON}, there still remains a
possibility that (at least) an essential feature of the top quark
condensate idea may eventually survive.
As a possible modification within the simplest version of the model 
we shall experiment with an idea 
to take the cutoff beyond the Planck scale.
Even if we were allowed to ignore the quantum gravity effects, 
however,  we cannot take the cutoff beyond
 the Landau pole of $U(1)_Y$ gauge coupling, 
 which actually yields an absolute minimum
 value of the top mass prediction $m_t \simeq 200 {\rm GeV}$.
 
On the other hand, if the standard 
 gauge groups are unified into a (``walking'')\footnote{
 In the context of renormalizability
 of the gauged NJL model, we shall use in this section
 ``walking'' for $A=c/b>1$ 
 (slowly running) instead of the usual definition of walking
$A\gg 1$ ({\it very} slowly running including non-running case).
 }  
 GUT,  we may take the cutoff to infinity 
 thanks to the renormalizability%
\cite{kn:KSY91,kn:KTY93,kn:Yama92,kn:Kras93,kn:KSTY94,kn:HKKN94}
  of the gauged NJL model with ``walking'' ($A>1)$ gauge coupling. 
  We shall consider this possibility 
  (``top mode walking GUT'')\cite{kn:ITY96} in which the top and Higgs mass 
 prediction is 
 controlled by the Pendleton-Ross (PR) infrared
 fixed point \cite{kn:PR81} at GUT scale and can naturally 
 lead to $m_t \simeq m_H \simeq 180  {\rm GeV}$.
\par

\subsection{The Model}

 Let us first explain the original version of the top quark condensate model
  (top mode standard model) proposed by MTY\cite{kn:MTY89a,kn:MTY89b}
  based on explicit four-fermion interactions. 
The model
consists of the standard three families of quarks and leptons 
with the standard $SU(3)_C \times SU(2)_L \times U(1)_Y$ gauge 
interactions 
but {\it without Higgs doublet}. Instead of the standard Higgs sector 
MTY introduced $SU(3)_C \times SU(2)_L \times U(1)_Y$-invariant 
{\it four-fermion interactions among quarks and leptons}, the origin of 
which is expected to be a new physics not specified at this moment. 
The new physics determines the ultraviolet (UV) scale (cutoff $\Lambda$) of
 the model, in contrast to the infrared (IR) scale (weak scale $F_{\pi}\simeq 
 250 \GeV$)
 determined by the mass of $W/Z$ bosons.

The explicit form of such four-fermion 
interactions reads: \cite{kn:MTY89a,kn:MTY89b}   
\begin{eqnarray}
  {\cal L}_{4f}
  &=&
   \bigg[
       G^{(1)} 
         (\bar\psi_L^i \psi_R^j)
         (\bar\psi_R^j \psi_L^i)
  \nonumber \\
  & & +G^{(2)} 
         (\bar\psi_L^i \psi_R^j)
         (i\tau_2)^{ik}(i\tau_2)^{jl}
         (\bar\psi_L^k \psi_R^l)
  \nonumber \\
  & &+ G^{(3)} 
         (\bar\psi_L^i \psi_R^j)
         (\tau_3)^{jk}
         (\bar\psi_R^k \psi_L^i)
   \bigg]
     +\hbox{h.c.},
\label{eq:(4.1)}
\end{eqnarray}
where $i,j,k,l$ are the weak isospin indices
 and $G^{(1)}$, $G^{(2)}$ and 
$G^{(3)}$ are the four-fermion coupling constants
among top and bottom quarks $\psi\equiv (t,b)$.
It is straightforward\cite{kn:MTY89a,kn:MTY89b} to include other families and
leptons into this form.

The symmetry structure (besides $SU(3)_C$) of the 
four-fermion interactions, $G^{(1)}$, $G^{(2)}$ and 
$G^{(3)}$, is 
$SU(2)_L \times SU(2)_R \times U(1)_V \times U(1)_A$, 
$SU(2)_L \times SU(2)_R \times U(1)_V $ and 
$SU(2)_L \times U(1)_Y \times U(1)_V \times U(1)_A$, respectively. 
The $G^{(2)}$ term is vital to the mass of
the bottom quark in this model.\cite{kn:MTY89a,kn:MTY89b}
In the absence of the $G^{(2)}$-term, (\ref{eq:(4.1)}) possesses a 
$U(1)_A$ symmetry 
which is explicitly broken only by the color anomaly and plays the role
of the Peccei-Quinn symmetry.\cite{kn:MTY89b}

Let us disregard the $G^{(2)}$ term for the moment, in which case
the MTY Lagrangian (\ref{eq:(4.1)}) simply reads
\begin{equation}
  {\cal L}_{4f}
   =G_t (\bar\psi_L t_R)^2 + G_b (\bar\psi_L b_R)^2 + h.c.,
\label{eq:tb-4fermi}
\end{equation}
with $G_t \equiv G^{(1)}+G^{(3)}$ and $G_b \equiv G^{(1)}-G^{(3)}$.
This MTY Lagrangian with $G_b=0$ was the starting point of
BHL\cite{kn:BHL90}, but setting $G_b=0$ overlooks an important
aspect of the top quark condensate, as we will see in the following. 

\subsection{Why $m_t \gg m_{b,c,\cdot\cdot\cdot}$?}

We now explain one of the key points of the model, i.e.,
explicit dynamics  which gives rise to a large isospin
violation in the condensate $\VEV{\bar tt}\gg \VEV{\bar bb}$
 ($m_t \gg m_b$), or more generally, naturally explains why
 only the top quark has a very large mass.
  MTY found\cite{kn:MTY89a,kn:MTY89b} that critical phenomenon, or
  theory having nontrivial UV fixed point with
  large anomalous dimension, is actually such a dynamics, based on the 
 \SxSB solution
of the ladder SD equation for the gauged NJL model. Such an amplification
of symmetry violation was already explained in NJL model and gauged NJL model
in the previous sections.

For the $SU(3)_c\times SU(2)_L\times U(1)_Y$-gauged NJL model, 
the ladder SD equation becomes simpler in the large $N_c$ limit: 
Rainbow diagrams of
the $SU(2)_L\times U(1)_Y$ gauge boson lines are suppressed compared with
those of the QCD gluon lines. 
 Thus we consider 
 the ladder SD equation with  QCD coupling 
 and four-fermion coupling (\ref{eq:tb-4fermi}).
Without $G^{(2)}$ term, the
top and bottom quarks satisfy decoupled SD equations,
each equation being the same form as (\ref{eq:(4.4)}) with different 
four-fermion couplings $g_t \not = g_b $ $(g^{(3)} \not = 0 )$.
(We can easily find a solution for the SD equation with the $G^{(2)}$ 
term.)\cite{kn:MTY89b}
For simplicity, we first consider the {\it non-running} QCD coupling ,
in which case (\ref{eq:(4.4)}) is reduced to (\ref{eq:(4.2)}).
 Then there exists a critical line (\ref{eq:(2.17)}) around which 
the dynamical mass $m$ in (\ref{scalinggNJL})
sharply rises from zero to order $O(\Lambda)$
as we move away from the critical coupling $g=g^*(\lambda)$. 
In view of the critical line and the critical
behavior (\ref{scalinggNJL}),
MTY\cite{kn:MTY89a,kn:MTY89b} indeed found  
{\it amplified
 isospin symmetry violation} for a small ({\it however small}) violation
 in the coupling constants. 
Thus we have an S$\chi$SB
solution with {\it maximal isospin violation}, $m_t \not =0$ and
 $m_b=0$,
when
\begin{equation}
 g_t> g^*={1\over 4}(1+\omega)^2 >g_b
\label{eq:tbsplit}
\end{equation}
($g_t$ is above the critical line and $g_b$ is below
it). As already mentioned, we need {\it not} to set $G_b =0$ 
in the four-fermion
 interactions (\ref{eq:tb-4fermi}) to obtain $m_b=0$. 
 Thus, even if we assume that all the dimensionless couplings are 
 $O(1)$,  the 
 {\it critical phenomenon}
 naturally explains why only the top quark can have a large mass,
 or more properly, {\it why other fermions can have very small masses}: 
 $m_t \gg m_{b,c,\cdot\cdot\cdot}$. It is indeed realized if
  only the top quark coupling is above the
 critical coupling, while all others below it: $g_t >g^*>
 g_{b,c,\cdot\cdot\cdot}$$ \Longrightarrow m_t \ne 0, m_{b,c,\cdot\cdot\cdot}=0$.
Note that other couplings do {\it not} need to be zero nor {\it very} small.

In the case of running gauge coupling  (\ref{eq:(4.4)}),
we have already seen that 
essentially the same critical
phenomenon takes place through the presence of 
the ``critical line'' (\ref{eq:``criticalline''}): 
We again have an S$\chi$SB solution with
{\it maximal isospin violation}, $m_t \ne 0 , m_b=0$ (apart from 
$m_{QCD}\ll m_t$)
, under a condition 
similar to (\ref{eq:tbsplit}); 
$
   g_t> g^* (\simeq 1-2\lambda_{\Lambda}) > g_b.
$

\subsection{Top Quark Mass Prediction}

Now we come to the central part of the model, namely,
relating the dynamical mass of the condensed fermion
 (top quark) to the mass of $W/Z$ bosons.
 
The top quark condensate $\VEV{\bar tt}$ indeed yields a standard gauge
 symmetry breaking pattern
$SU(2)_L \times U(1)_Y \rightarrow U(1)_{em}$ 
to feed the mass of W and Z bosons.
Actually, the mass of $W$ and $Z$ bosons in the top quark condensate
 is generated via dynamical Higgs mechanism as 
in the technicolor, (\ref{eq:W/Zmass}),
where $F_{\pi} (\simeq 250{\rm GeV})$ determine the IR scale of the model,
this time the top quark mass.

\subsubsection{SD Equation plus PS Formula (MTY)}

The decay constants of these composite NG bosons $F_\pi$ may be calculated 
through the formula (\ref{fpiinteg}) which are written in terms of 
the fermion mass function and the amputated
BS amplitude. The BS amplitude is a solution of the BS equation 
 which must be solved consistently with 
the SD equation for the fermion propagator.\cite{kn:ABKMN90}
Instead, here we use  the PS formula 
(\ref{PS-QCD})\cite{kn:PS79} for simplicity,
which was generalized by MTY\cite{kn:MTY89a} to 
the $SU(2)$-asymmetric case, 
$m_t \not = m_b$ and $m_{t,b} \not =0$:
\begin{eqnarray}
  F_{\pi^\pm}^2 
    &=& 
      {N_c\over 8\pi^2} \int_0^{\Lambda^2}dxx\cdot
    \nonumber \\
    & & \cdot
  {\DISP (\Sigma_t^2+\Sigma_b^2)-{x\over4}(\Sigma_t^2+\Sigma_b^2)^\prime
   +{x\over2}(\Sigma_t^2-\Sigma_b^2)
   \left[
      {\DISP 1+(\Sigma_t^2)^\prime \over \DISP x+\Sigma_t^2}
     -{\DISP 1+(\Sigma_b^2)^\prime \over \DISP x+\Sigma_b^2}
   \right]
   \over
   \DISP (x+\Sigma_t^2)(x+\Sigma_b^2)
  },
  \label{eq:(4.6)}\\
  F_{\pi^0}^2 
    &=& {N_c\over 8\pi^2} \int_0^{\Lambda^2}dxx\cdot
    \left[
      {\DISP \Sigma_t^2-{x\over 4}(\Sigma_t^2)^\prime
       \over
       \DISP (x+\Sigma_t^2)^2
      }
     +{\DISP \Sigma_b^2-{x\over 4}(\Sigma_b^2)^\prime
       \over
       \DISP (x+\Sigma_b^2)^2
      }
    \right].
 \label{eq:(4.7)}
\end{eqnarray}

Let us consider the extreme case, the maximal isospin
 violation mentioned above, $\Sigma_t(p^2) \ne 0$ and $\Sigma_b(p^2)=0$.
We further take a ``toy'' case switching off the gauge interactions: 
$\Sigma_t (p^2)\equiv {\rm const.}=m_t$ (pure NJL limit).
Then (\ref{eq:(4.6)}) and
(\ref{eq:(4.7)}) are both {\it logarithmically divergent} at
 $\Lambda/ m_t \rightarrow \infty$ 
{\it with the same coefficient}:
\begin{eqnarray}
  F_{\pi^\pm}^2 
  &=& {N_c \over 8\pi^2}m_t^2
  \left[
    \ln{\Lambda^2 \over m_t^2} +{1 \over 2}\right],
  \label{eq:F-NJL1}\\
  F_{\pi^0}^2 
  &=& {N_c \over 8\pi^2}m_t^2 \ln{\Lambda^2 \over m_t^2}.
\label{eq:F-NJL2}
\end{eqnarray}
Now, we could predict $m_t$ by fixing $F_{\pi^\pm}
\simeq 250{\rm GeV}$ 
so as to have a correct $m_W$ through
(\ref{eq:W/Zmass}). Actually,  (\ref{eq:F-NJL1}) determines $m_t$ as a
 {\it decreasing function of cutoff $\Lambda$}.
  The largest physically sensible $\Lambda$ (new physics scale) would be
 the Planck scale $\Lambda\simeq 10^{19}\GeV$ at which we have a 
minimum value prediction
  $m_t \simeq 145\GeV$. If we take the limit 
 $\Lambda \rightarrow \infty$, we would have $m_t \rightarrow 0$,
 which is nothing but triviality
 (no interaction) of the pure NJL model: $y_t \equiv\sqrt{2}
  m_t/F_\pi \rightarrow 0$
 at $\Lambda \rightarrow \infty$. 

One might naively expect a disastrous weak isospin violation for the maximal 
isospin-violating dynamical mass, $m_t \ne 0$ and $m_b =0$.
 However, for 
 $\Lambda\gg m_t$, (\ref{eq:F-NJL1}) and (\ref{eq:F-NJL2}) yield 
$
F_{\pi^\pm} \simeq F_{\pi^0},
$
or
\begin{equation}
  \delta\rho 
  \equiv {F_{\pi^\pm}^2- F_{\pi^0}^2 \over  F_{\pi^{\pm}}^2}
   ={N_c m_t^2 \over 16\pi^2 F_{\pi^{\pm}}^2} 
   \simeq {1 \over 2 \ln{{\Lambda^2 \over m_t^2}}}
   \ll 1.
\label{eq:(4.8)}
\end{equation}
Then the problem of weak isospin 
 relation can in principle be solved {\it without custodial symmetry}.
Actually, the isospin violation 
$
F_{\pi^\pm} \ne F_{\pi^0}
$
in (\ref{eq:(4.6)}) and (\ref{eq:(4.7)})
solely comes from the different propagators having different 
$\Sigma (p^2)$, essentially the IR quantity, which becomes less important 
for $\Lambda \gg m$, since the integral is UV dominant.
This is the essence of the ``dynamical mechanism'' of MTY to
save the isospin relation $\rho \simeq 1$ without custodial 
symmetry.

Now in the gauged NJL model, QCD plus four-fermion interaction
 (\ref{eq:tb-4fermi}), essentially the same mechanism  
 as the above is operative. Based on the very slowly damping 
 solution of the ladder SD equation (\ref{eq:(4.5)}) and 
 the PS formulas (\ref{eq:(4.6)})-(\ref{eq:(4.7)}), MTY\cite{kn:MTY89a,kn:MTY89b} predicted
$m_t$ and $\delta \rho$ as the {\it decreasing function of
 cutoff $\Lambda$}.
For the Planck scale cutoff $\Lambda\simeq 10^{19} \GeV$,
we have\cite{kn:MTY89a,kn:MTY89b}\footnote{
One may substitute into (\ref{eq:(4.5)}) the {\it numerical} solution
( instead of the analytical one
(\ref{eq:(4.5)})) of the ladder SD equation 
(\ref{eq:(4.4)}), the result being the same as 
(\ref{eq:(4.15)}).\cite{kn:King90}
}\addtocounter{footnote}{1} 
\begin{equation}
        m_t \simeq 250 \GeV,
\label{eq:(4.15)}
\end{equation}
\begin{equation}
        \delta \rho \simeq 0.02 \ll 1.          
\label{eq:(4.16)}
\end{equation}
This is compared with the pure NJL case $m_t \simeq 145\GeV$: The QCD
corrections are quantitatively rather significant
 (Presence of the gauge
coupling also changes the qualitative feature of the theory
from a nonrenormalizable/trivial theory into a renormalizable/nontrivial
one.)\cite{kn:KSY91,kn:KTY93,kn:Yama92,kn:Kras93,kn:KSTY94,kn:HKKN94}

It will be more convenient to write an {\it analytical}
 expression for $F_{\pi}$.
  Neglecting the derivative terms with $\Sigma_t (x)'$ and using
  (\ref{eq:(4.5)}), we may approximate (\ref{eq:(4.6)}) as\cite{kn:Yama92} 
\begin{eqnarray}
F_{\pi}^2
  &\simeq&  {N_c\over 8\pi^2}\int_{m_t^2}^{\Lambda^2}dx{\Sigma_t^2 \over x}
  \nonumber \\
  &\simeq&  {N_c m_t^2 \over 16\pi^2}{A \over A-1}
               {(\lambda(m_t^2))^{A-1} - (\lambda(\Lambda^2))^{A-1}
                  \over   
                     (\lambda(m_t^2))^{A}}.
\label{eq:marciano}
\end{eqnarray}
This analytic expression was first obtained by Marciano\cite{kn:Marc89} 
in the case of $A=8/7$ ($ N_f=6$), which actually reproduces the MTY 
prediction (\ref{eq:(4.15)}).

\subsubsection{RG Equation plus Compositeness Condition (BHL)}

Now, we explain the BHL formulation\cite{kn:BHL90}
 of the top quark condensate, which is based on the RG equation combined with
 the compositeness condition. 
BHL start with the SM Lagrangian
which includes explicit Higgs field at the Lagrangian level:
\begin{equation}
{\cal L}_{SM} = -y_t (\bar \psi_L^i t_R \phi _i + \mbox{h.c.})
 +  \left(D_\mu \phi ^\dagger\right)\left(D^\mu \phi \right)
  -   m_H^2 \phi ^\dagger\phi 
   -  \lambda_4 \left(\phi ^\dagger\phi \right)^2,
\label{eq:top51} 
\end{equation}
where $y_t$ and $\lambda_4$ are Yukawa coupling of the top quark and 
quartic interaction of the Higgs, respectively.
BHL imposed ``compositeness condition'' on $y_t$ and $\lambda_4$ in such a
way that (\ref{eq:top51}) becomes the {\it MTY Lagrangian} (\ref{eq:tb-4fermi})
(with $G_b=0$):
\begin{equation}
  \dfrac{1}{y_t^2}\rightarrow 0, \quad
   \dfrac{\lambda _4}{y_t^4}\rightarrow 0 \qquad 
    \mbox{as }    \mu \rightarrow \Lambda,
\label{eq:top53}
\end{equation}
where $\mu$ is the renormalization point above which the composite dynamics
are integrated out to yield an effective theory (\ref{eq:top51}).
Thus the compositeness condition implies divergence 
at $\mu = \Lambda $ of both the Yukawa
coupling of the top quark and the quartic interaction of the Higgs. 

Now, in the one-loop RG equation, the beta function of $y_t$ is given by
\begin{equation}
  \beta(y_t) = \dfrac{y_t^3}{(4\pi )^2}
              \left(N_c + \dfrac{3}{2}\right)
             - \dfrac{y_t}{(4\pi )^2}
                \left(3\frac{N_c^2-1}{N_c}g_3^2
                +\frac{9}{4}g_2^2-\frac{17}{12}g_1^2\right),
\label{eq:top61}
\end{equation}
where $g_1$, $g_2$ and $g_3$ are the gauge couplings of $U(1)_Y$, $SU(2)_L$
and $SU(3)_C$, respectively. BHL solved the RG equation for the
beta function (\ref{eq:top61}) combined with 
the compositeness condition (\ref{eq:top53}) as a boundary condition
at $\mu =\Lambda$.\\

\subsubsection{BHL versus MTY}

Let us 
first demonstrate\cite{kn:Yama92} 
that {\it in the large $N_c$ limit} BHL formulation\cite{kn:BHL90} 
is {\it equivalent} to that of MTY\cite{kn:MTY89a,kn:MTY89b}, both based on
the same MTY Lagrangian (\ref{eq:tb-4fermi}).
In the $N_c\rightarrow \infty$ limit for (\ref{eq:top61}),
we may neglect the factor $3/2$ in the first term 
(composite Higgs loop effects) and $g_2^2$ and $g_1^2$
in the second term (electroweak gauge boson loops), which
corresponds to the similar neglection of $1/N_c$ sub-leading
effects in the ladder SD equation in the MTY approach. 
Then (\ref{eq:top61}) becomes simply:
\begin{equation}
  \dfrac{d y_t}{d \mu}= \beta(y_t) = N_c \dfrac{y_t^3}{(4\pi )^2}  
                       - \dfrac{3N_c y_t g_3^2}{(4\pi )^2}.
\label{eq:top62}
\end{equation}
Within the same approximation the beta function of the 
QCD gauge coupling reads
\begin{equation}
  \dfrac{d g_3}{d \mu} = \beta (g_3) = -\frac{1}{A} 
  \dfrac{3N_c g_3^3}{(4\pi )^2}.
\label{eq:top63}
\end{equation}
Solving (\ref{eq:top62}) and (\ref{eq:top63}) by imposing 
the compositeness condition at $\mu =\Lambda $, we arrive at\cite{kn:Yama92}
\begin{equation}
  y_t^2(\mu ) = 
     \dfrac{2(4\pi )^2}{N_c} \frac{A-1}{A} 
     \dfrac{\left(\lambda (\mu^2 )\right)^A}
     {\left(\lambda (\mu^2 )\right)^{A-1}
     - \left(\lambda (\Lambda^2 )\right)^{A-1}}.
\end{equation}
Noting the usual relation $m_t^2=\frac{1}{2}y_t^2(m_t)\:v^2\:
(v= F_{\pi})$,
we obtain 
\begin{equation}
 \dfrac{m_t^2}{F_{\pi}^2}
   = \dfrac{y_t^2(m_t)}{2}=
     \dfrac{(4\pi )^2}{N_c} \frac{A-1}{A} 
     \dfrac{\left(\lambda (m_t^2 )\right)^A}
     {\left(\lambda (m_t^2 )\right)^{A-1}
     -\left(\lambda (\Lambda^2 )\right)^{A-1}}.
\label{BHLyukawa}
\end{equation}
This is precisely the same formula as (\ref{eq:marciano}) obtained in the
 MTY approach based on the SD 
 equation and the PS formula.\footnote{
 Alternatively, we may define $F_{\pi}^2(\mu^2) \equiv 2m_t^2/y_t^2(\mu)$
  which coincides 
 with the integral (\ref{eq:marciano}) with the IR end $m_t^2$ simply replaced 
  by $\mu^2$. Then the compositeness condition (\ref{eq:top53}) reads
  $F_{\pi}^2(\mu^2=\Lambda^2)=0$ (no kinetic term of the 
  Higgs).
  }
  Thus we have established
  \begin{equation}
  {\rm BHL} (\frac{1}{N_c}\: {\rm leading}) = {\rm MTY}.
  \end{equation}
 
Having established equivalence between MTY and BHL in the
large $N_c$ limit, we now
comment on the relation between them in more details.
 Note that MTY formulation is based on the nonperturbative picture,
 ladder SD equation and PS formula, which is valid at
 $1/N_c$ leading order, or the NJL bubble sum 
 with ladder-type QCD corrections
 (essentially the leading log summation).
 MTY extrapolated this $1/N_c$ leading picture all the
 way down to the low energy region where the sub-leading effects
 may become important. 
 
 On the other hand, BHL is crucially based on  
 the perturbative picture, one-loop RG equation, which can
 easily accommodate $1/N_c$ sub-leading effects 
 in (\ref{eq:top61}) such as 
 the loop effects of composite Higgs and electroweak gauge bosons.
 However, BHL formalism must necessarily be combined with the 
 compositeness condition (\ref{eq:top53}). The compositeness condition
  is obviously inconsistent with the perturbation and
 is a purely nonperturbative concept based on the same $1/N_c$ leading 
 NJL bubble sum as in the MTY formalism.
 Thus the BHL perturbative picture breaks down at high energy 
 near the compositeness scale $\Lambda$ where
 the couplings $y_t$ and $\lambda_4$ blow up as required by the 
 compositeness condition.

 So there must be a certain
 matching scale $\Lambda_{\rm Matching}$  such that 
 the perturbative picture (BHL) is valid for $\mu<\Lambda_{\rm Matching}$, 
 while 
 only the nonperturbative picture (MTY) 
 becomes consistent for $\mu >\Lambda_{\rm Matching}$.\footnote{
 Of course, the $1/N_c$ leading picture 
 might be subject to ambiguity such as the possible higher dimensional 
 operators, cutoff procedures, etc., all related to the nonrenormalizability
 of the NJL model.\cite{kn:Suzu90} 
 These problems will be conceptually solved and 
 phenomenologically tamed, when coupled
 to the (``walking'' ($A>1$))
 gauge interactions (renormalizability of the
 gauged NJL 
 model)\cite{kn:KSY91,kn:KTY93,kn:Yama92,kn:Kras93,kn:KSTY94,kn:HKKN94} to be
 discussed later. Here we just comment that even if
 there might be such an ambiguity, the $1/N_c$ picture (MTY)
 is the only 
 consistent way to realize the compositeness condition as was done by
 the BHL paper itself.  
 }
 Such  a point may be defined by the energy region where the two-loop
 contributions dominate over the one-loop ones. 
 However, 
 thanks to the presence of a {\it quasi-infrared fixed point}\cite{kn:Hill81},
 BHL prediction is {\it numerically}
  quite stable against ambiguity at high energy region,
 namely, rather independent of whether this high energy region is 
 replaced by MTY or something else.
 Then we expect $m_t \simeq m_t ({\rm BHL}) =
 \frac{1}{\sqrt{2}} y_t(\mu=m_t) v \simeq \frac{1}{\sqrt{2}} \bar y_t v$ 
 within 1-2 \%, where 
$\bar y_t$ is the quasi-infrared fixed point given by $\beta(\bar y_t)=0$
in (\ref{eq:top61}). The composite Higgs loop changes ${\bar y_t}^2$ by
roughly the factor $N_c/(N_c+3/2)=2/3$ compared with the MTY value, i.e.,
$250 {\rm GeV}\rightarrow 250\times \sqrt{2/3}=204 {\rm GeV}$, 
while the electroweak gauge boson loop with opposite sign pulls it 
back a little bit to a higher value.
The BHL value\cite{kn:BHL90} is then given by
\begin{equation}
  m_t=218\pm3 {\rm GeV}, \qquad \mbox{at } \Lambda \simeq 10^{19}\GeV.
\end{equation}

The Higgs boson was predicted as a $\bar t t$ bound state 
with a mass $M_H \simeq 2 m_t$ \cite{kn:MTY89a,kn:MTY89b,kn:Namb89} 
based on the pure NJL model calculation\cite{kn:NJL61}.
Its mass was also calculated by BHL\cite{kn:BHL90} 
through the full RG equation of $\lambda_4$, the result being
\begin{equation}
  M_H = 239\pm3 {\rm GeV} \qquad (\frac{M_H}{m_t}\simeq 1.1)
   \qquad \mbox{at } \Lambda \simeq 10^{19}\GeV.
\end{equation}
If we take only the $1/N_c$ leading terms, we would have the mass ratio 
$M_H/m_t \simeq \sqrt{2}$, which was also obtained through the 
ladder SD equation.\cite{kn:STY90}

\subsection{Top Mode Walking GUT}

As we have seen, the top quark condensate naturally explains, through
the critical phenomenon, why only the top
quark mass is much larger than that of
other quarks and leptons: $m_t \gg m_{b,c,\cdot\cdot\cdot}$.
It further predicts the top mass on the order of weak scale. However, the 
predicted mass $220 {\rm GeV}$ is somewhat larger than the mass of the 
recently discovered top quark, $176 {\rm GeV}\pm 13 {\rm GeV}$ (CDF) and
$199 +38/-36 {\rm GeV}$ (D0)\cite{kn:TEVATRON}. 
Here we shall discuss a possible remedy of this
problem within the simplest model based on the MTY Lagrangian 
(\ref{eq:tb-4fermi}).\cite{kn:ITY96}

\subsubsection{Landau Pole Scenario}

First we recall that the top mass prediction is a {\it decreasing} function of 
the cutoff $\Lambda$. Then the simplest way to reduce the top mass would be
to raise the cutoff as much as possible. Let us assume
 that quantum gravity effects would not change drastically 
 the physics described by the low energy  theory without gravity.
 Then we may raise the cutoff $\Lambda$ beyond the Planck scale up to
 the Landau pole $\Lambda \simeq 10^{41} {\rm GeV}$ 
 where the $U(1)_Y$ gauge coupling $g_1$ diverges and the SM description 
 itself stops to be self-consistent.
In such a case the top and Higgs mass prediction becomes:
\begin{equation}
m_t \simeq 200 {\rm GeV}, \qquad  M_H \simeq 209 {\rm GeV}
\qquad \mbox{at } \Lambda\simeq 10^{41}\GeV
\end{equation}
which is the absolute minimum value of the prediction
within the simplest version of the top quark condensate.

If it is really the case, it would imply composite $U(1)_Y$ gauge
boson and composite Higgs generated {\it at once by the same dynamics}, 
since the Landau pole then may be regarded as a BHL compositeness condition 
also for the vector bound state as well as the composite Higgs.
Actually, we can formulate the BHL compositeness condition for
vector-type four-fermion interactions
 (Thirring-type four-fermion theory) 
as a {\it necessary condition} for the 
formation of a vector bound state. The possibility that both the
Higgs and $U(1)_Y$ gauge boson can be composites by the same dynamics
may be illustrated by an explicit model, the Thirring model in $D (2<D<4)$
dimensions. Reformulated as a gauge theory through hidden local 
symmetry\cite{kn:BKY88}, the Thirring model was shown
to have the dynamical mass 
generation, which implies that a composite Higgs 
and a composite gauge boson are generated at the same time.\cite{kn:IKSY95}

At any rate, the prediction of this scenario $m_t \simeq 200 {\rm GeV}$
still seems to be a little bit higher than the experimental
value, although the situation is not very conclusive yet.
Then we shall consider another possibility, namely, taking
 the cutoff to infinity: $\Lambda\rightarrow \infty$. 
In order to do this we should first recall the previous discussions 
on the renormalizability
of the gauged NJL model with ``walking'' gauge 
coupling 
($A>1$).\cite{kn:KSY91,kn:KTY93,kn:Yama92,kn:Kras93,kn:KSTY94,kn:HKKN94}

\subsubsection{More on Renormalizability of Gauged NJL Model}

This phenomenon was first pointed out by Kondo, Shuto and 
Yamawaki\cite{kn:KSY91}
through the convergence of $F_{\pi}$ in the PS formula 
for the solution of the SD equation (\ref{eq:(4.5)}) in the
four-fermion theory plus QCD. 
 Contrary to the logarithmic divergence of (\ref{eq:F-NJL1})
in the pure NJL model, it was emphasized that 
for $A>1$ we have a {\it convergent} 
integral for $F_{\pi}$ and hence 
a nontrivial (interacting)
 theory with finite effective Yukawa coupling 
$y_t\equiv \sqrt{2} m_t/F_{\pi}\ne 0$ in the continuum limit:
 Namely, the presence of ``walking'' ($A>1$) gauge interaction 
changes the trivial/nonrenormalizable theory (pure NJL model) into a 
nontrivial/renormalizable theory (gauged NJL model).\cite{kn:KSY91}

The analytical 
expression of the effective Yukawa coupling is already given by 
(\ref{eq:marciano}) (MTY), which is equivalent to  
 (\ref{BHLyukawa}) obtained as a solution of 
the RG equation with a compositeness condition at $1/N_c$ leading (BHL).
 From this expression it was again noted\cite{kn:Yama92} 
that {\it iff $A>1$} (``walking'' gauge coupling with $N_c\sim N_f\gg 1$), 
then the effective Yukawa coupling remains finite, 
$y_t >0$, in the continuum limit $\Lambda \rightarrow \infty$.
This is in sharp contrast to the triviality of the
pure NJL model in which $y_t \rightarrow 0$ in the continuum
limit as was mentioned earlier. 

It was further pointed out
by Kondo, Tanabashi and Yamawaki\cite{kn:KTY93}
that this renormalizability is equivalent to existence of
a PR infrared fixed point\cite{kn:PR81} for the gauged
Yukawa model. The PR fixed point is given by the solution of
$\dfrac{d (y_t/g_3)}{d \mu}=0$ with 
(\ref{eq:top62}) and (\ref{eq:top63}):
\begin{equation}
y_t^2 
      =\frac{(4\pi)^2}{N_c}\frac{A-1}{A}\lambda,
\label{PR}
\end{equation}
where $\lambda=3 C_2(\mbox{\boldmath$F$}) g_3^2/(4\pi)^2$. Similar argument was recently 
developed more systematically by Harada, Kikukawa, Kugo and 
Nakano.\cite{kn:HKKN94}.

As to the non-running (standing) case ($A \rightarrow \infty$), 
the integral for 
$F_{\pi}^2$ is more rapidly convergent, since $\Sigma(p^2)$ is power
damping, (\ref{eq:(4.3)}), instead of logarithmic damping. 
In this case the renormalization
procedure was performed explicitly by Kondo, Tanabashi and 
Yamawaki \cite{kn:KTY93} through the effective 
potential in the ladder approximation  as was already explained in Section 5.

\subsubsection{Top Mode Walking GUT}
In view of the renormalizability
of the gauged NJL model with ``walking'' gauge coupling, we may take 
the $\Lambda \rightarrow \infty$ limit of the
top quark condensate.
However, in the realistic case we actually have the 
 $U(1)_Y$ gauge coupling which, as it stands, grows at high energy
 to blow up at Landau pole and hence 
 invalidates the above arguments of the renormalizability. Thus, in order to
apply the above arguments to the top quark condensate, 
we must remove the $U(1)_Y$ gauge interaction in such a way
as to unify it into a GUT with ``walking'' coupling ($A>1$)
beyond GUT scale. 
Then the renormalizability requires that the GUT coupling at GUT
scale should be determined by the PR infrared fixed point.\cite{kn:ITY96}

For a simple-minded GUT with $SU(N)$  group, the PR fixed point
takes the form similar to (\ref{PR}):
\begin{eqnarray}
y_t^2 (\Lambda_{\rm GUT}) 
&=& \frac{3 C_2(\mbox{\boldmath$F$})}{N}\frac{A-1}{A} 
g_{\rm GUT}^2 (\Lambda_{\rm GUT})
          \simeq \frac{3}{2} g_{\rm GUT}^2 (\Lambda_{\rm GUT}),\nonumber \\
\lambda_4 (\Lambda_{\rm GUT})
&=&\frac{6 C_2(\mbox{\boldmath$F$})}{N}\frac{(A-1)^2}{A(2A-1)} 
g_{\rm GUT}^2 (\Lambda_{\rm GUT})
          \simeq \frac{3}{2} g_{\rm GUT}^2 (\Lambda_{\rm GUT}),
\end{eqnarray}
where we assumed $N\gg 1$ and $A\gg 1$ ($N_f\sim N\gg 1$) 
for simplicity. Then the 
top Yukawa coupling at GUT scale is essentially determined by the GUT 
coupling at GUT scale up to some numerical factor depending on the GUT
group and the representations of particle contents. Using 
``effective GUT coupling'' including such possible numerical factors, 
we may perform the BHL full RG equation analysis 
for $\mu <\Lambda_{\rm GUT} \simeq 10^{15} {\rm GeV}$
with the boundary condition of the above PR fixed
point at GUT scale.

For typical values of the effective GUT coupling $\alpha_{\rm GUT}\equiv 
g_{\rm GUT}^2/4\pi=1/40,\: 1/50$ and 1/60,
prediction of the top and Higgs masses reads:
\begin{equation}
(m_t, M_H)\simeq (189, 193),\: (183, 183),\: 
(177,173)\qquad {\rm GeV},
\end{equation}
respectively. Note that these PR fixed point values at GUT scale 
are somewhat smaller than the coupling values at GUT scale 
which focus on the quasi-infrared fixed point in the low energy region.
Thus the prediction is a little bit away from the quasi-infrared fixed point. 
This would be the simplest extension of the top quark condensate 
consistent with the recent experiment on the top quark mass.

\section{Conclusion}
We have discussed a variety of tightly bound composite Higgs models,
walking technicolor, strong ETC technicolor and top quark condensate, 
based on
the gauged NJL model as the explicit dynamics with large anomalous 
dimension and fixed point. Universal feature
of this type of dynamics is the amplification of the symmetry violation
due to the large anomalous dimension or the fine-tuning of the coupling near
the critical point (UV fixed point). An extreme case $\gamma_m \simeq 2$ 
yields maximal symmetry violation, which was used to predict an exceptionally
large mass of the top quark, even if the top coupling is on the same order
as those of other quarks and leptons. The fine-tuning of the gauged NJL
model corresponds to the renormalization which leads to
the existence of the renormalizable and nontrivial
continuum theory in contrast to the pure NJL model.   
Although the situation about top quark mass is still not yet conclusive,
we hope that at least essence of the idea of the top quark condensate
may eventually survive in the sense that the {\it origin of mass}
is deeply related to the top quark mass. 

\medskip
{\bf Acknowledgements}\\

We would like to thank Iwana Inukai and Masaharu Tanabashi for
 collaboration and discussions on the recent results presented in this 
lecture. This work was supported in part
 by the Sumitomo Foundation and a Grant-in Aid for
    Scientific Research from the Ministry of Education, Science and Culture 
    (No. 05640339).


\begin{thebibliography}{99}
\bibitem{kn:NJL61} 
  Y. Nambu and G. Jona-Lasinio,
  Phys. Rev. {\bf 122} (1961) 345
\bibitem{kn:Wein76} 
  S. Weinberg, 
  Phys. Rev. {\bf D13} (1976) 974; {\bf D19} (1979) 1277;
  L. Susskind, 
  Phys. Rev. {\bf D20} (1979) 2619.
\bibitem{kn:FS81}  
  For a review,
  E. Farhi and L. Susskind, 
  Phys. Rep. {\bf 74} (1981) 277.
\bibitem{kn:Hold85}
  B. Holdom, 
  Phys. Lett. {\bf B150} (1985) 301.
\bibitem{kn:YBM86} 
  K. Yamawaki, M. Bando and K. Matumoto, 
  Phys. Rev. Lett. {\bf 56} (1986) 1335;
  T. Akiba and T. Yanagida, 
  Phys. Lett. {\bf B169} (1986) 432;
  T. Appelquist and L.C.R. Wijewardhana, 
  Phys. Rev. {\bf D36} (1987) 568.
\bibitem{kn:MY89} 
  V.A. Miransky and K. Yamawaki,
  Mod. Phys. Lett. {\bf A4} (1989) 129.
\bibitem{kn:ATEW89}
  T. Appelquist, T. Takeuchi, M. Einhorn and L.C.R. Wijewardhana,
  Phys. Lett. {\bf B220} (1989) 223;
  K. Matumoto,
  Prog. Theor. Phys. {\bf 81} (1989) 277. 
\bibitem{kn:MTY89a} 
  V.A. Miransky, M. Tanabashi and K. Yamawaki,
  Phys. Lett. {\bf B221} (1989)177.
\bibitem{kn:MTY89b} 
  V.A. Miransky, M. Tanabashi and K. Yamawaki, 
  Mod. Phys. Lett. {\bf A4} (1989) 1043.
\bibitem{kn:Namb89} 
  Y. Nambu, 
  Chicago preprint EFI 89-08 (Feb., 1989).
\bibitem{kn:Marc89} 
  W.J. Marciano, 
  Phys. Rev. Lett. {\bf 62} (1989) 2793; 
  Phys. Rev. {\bf D41} (1990) 219.
\bibitem{kn:BHL90}  
  W.A. Bardeen, C.T. Hill and M. Lindner, 
  Phys. Rev. {\bf D41} (1990) 1647.
\bibitem{kn:BLL86} 
  W.A. Bardeen, C.N. Leung and S.T. Love, 
  Phys. Rev. Lett. {\bf 56} (1986) 1230;
  C.N. Leung, S.T. Love and W.A. Bardeen,
  Nucl. Phys. {\bf B273} (1986) 649.
\bibitem{kn:KMY89} 
  K.-I. Kondo, H. Mino and K. Yamawaki,
  Phys. Rev. {\bf D39} (1989) 2430; K. Yamawaki,
  {\it in Proc. Johns Hopkins Workshop on 
  Current Problems in Particle Theory 12, Baltimore, 1988,}
  eds. G. Domokos and S. Kovesi-Domokos 
  (World Scientific, Singapore, 1988);
  T. Appelquist, M. Soldate, T. Takeuchi and L.C.R. Wijewardhana, {\it ibid}.
\bibitem{kn:KSY91}
  K.-I. Kondo, S. Shuto and K. Yamawaki,
  Mod. Phys. Lett. {\bf A6} (1991) 3385.
\bibitem{kn:KTY93}
  K.-I. Kondo, M. Tanabashi and K. Yamawaki,
  Prog. Theor. Phys. {\bf 89} (1993) 1299. For an earlier version,
  Nagoya/Chiba preprint DPNU-91-20/Chiba-EP-53 (July, 1991), 
  Mod. Phys. Lett. {\bf A8} (1993) 2859. 
\bibitem{kn:Yama92}
  K. Yamawaki, {\it Proc. Workshop on Effective Field Theories of the
  Standard Model, Dobog\'ok\H\o, Hungary, August 22-26, 1991}, ed.
  U.-G. Meissner (World Scientific Pub. Co., Singapore, 1992) 307;
  M. Tanabashi, {\it Proc. Int. Workshop on Electroweak Symmetry Breaking,
  Hiroshima, Nov. 12-15, 1991,} eds. W.A. Bardeen, J. Kodaira and T. Muta
  (World Scientific Pub. Co., Singapore, 1992) 75.
\bibitem{kn:Kras93}
  N.V. Krasnikov,
  Mod. Phys. Lett. {\bf A8} (1993) 797. 
\bibitem{kn:KSTY94}
  K.-I. Kondo, A. Shibata, M. Tanabashi and K. Yamawaki,
  Prog. Theor. Phys. {\bf 91} (1994) 541; {\bf 93} (1995) 489. 
\bibitem{kn:HKKN94}
 M. Harada, Y. Kikukawa, T. Kugo and H. Nakano,
 Prog. Theor. Phys. {\bf 92} (1994) 1161. 
\bibitem{kn:Yama89}
 For example,  K. Yamawaki,
 {\it in Proc. of 1988 International Workshop on 
  New Trends in Strong Coupling Gauge Theories, Nagoya, Aug. 24-27, 1988,}
  eds. M. Bando, T. Muta and K. Yamawaki 
  (World Scientific Pub. Co., Singapore, 1989);
 {\it in Proc. 1989 Workshop on Dynamical Symmetry Breaking, Nagoya, 
  1989,} eds. T. Muta and K. Yamawaki (Nagoya University, 1990);
 {\it in Proc. 1990 International Workshop on Strong Coupling Gauge Theories
  and Beyond, Nagoya, 1990,} eds. T. Muta and K. Yamawaki
   (World Scientific Pub. Co., Singapore, 1991);\\
  M. Tanabashi, in {\it Proc. 1991 Nagoya Spring School on Dynamical 
  Symmetry Breaking, Nagoya, Japan} ed. K. Yamawaki
   (World Scientific Pub. Co., Singapore, 1992).
\bibitem{kn:Mira94}
  V.A. Miransky, 
  {\it Dynamical Symmetry Breaking in Quantum Field Theories}
  (World Scientific Pub. Co., Singapore, 1992).
\bibitem{kn:ITY96}
 I. Inukai, M. Tanabashi and K. Yamawaki,
 in preparation;
 K. Yamawaki, Nagoya preprint DPNU-96-09, to appear 
{\it in Proc. YKIS '95 ``From the Standard Model to Grand Unified 
Theories'', YITP, Kyoto University, Aug. 21-25, 1995}, ed. T. Kugo
(Suppl. Prog. Theor. Phys. 1996). 
\bibitem{kn:TEVATRON}
 F. Abe {\it et al.} (CDF), Phys. Rev. Lett. {\bf 74} (1995) 2626;
 S. Adachi {\it et al} (D0), Phys. Rev. Lett. {\bf 74} (1995) 2632.
\bibitem{kn:CW} 
  S. Coleman and E. Weinberg, 
  Phys. Rev. {\bf D7} (1973) 1888.
\bibitem{kn:BKY88}
  M. Bando, T. Kugo and K. Yamawaki,
  Phys. Rep. {\bf 164} (1988) 217.
\bibitem{kn:KTY90}
  K.-I. Kondo, M. Tanabashi and K. Yamawaki,
  {\it in Proc. 1989 Workshop on Dynamical Symmetry Breaking, Nagoya, 
  1989,} eds. T. Muta and K. Yamawaki (Nagoya University, 1990);
  D.E. Clague and G.G. Ross, Nucl Phys. {\bf B364} (1991) 43.  
\bibitem{kn:Lane74} 
  K. Lane, 
  Phys. Rev. D {\bf 10} (1974) 2605; 
  H.D. Politzer, 
  Nucl. Phys. {\bf B117} (1976) 397. 
\bibitem{kn:PS79} 
  H. Pagels and S. Stokar, 
  Phys. Rev.  {\bf D20} (1979) 2947.  
\bibitem{kn:Hold81}
  B. Holdom,
  Phys. Rev. {\bf D24} (1981) 1441.
\bibitem{kn:NNT91}
  Y. Nagoshi, K. Nakano and S. Tanaka,
  Prog. Theor. Phys. {\bf 85} (1991) 131;
\bibitem{kn:BKS91} 
  M. Bando, T. Kugo and K. Suehiro,
  Prog. Theor. Phys. {\bf 85} (1991) 1299. 
\bibitem{kn:KY90} 
  Y. Kikukawa and K. Yamawaki, 
  Phys. Lett. {\bf B234} (1989) 497.
\bibitem{kn:HKWY92}
  H.-J. He, Y.-P. Kuang, Q. Wang and Y.-P. Yi,
  Phys. Rev. {\bf D45} (1992) 4610.
\bibitem{kn:RWP90} 
  For a review, 
  B. Rosenstein, B.J. Warr and S.H. Park,
  Phys. Rep. {\bf 205} (1991) 59.
\bibitem{kn:Suzu90}
  M. Suzuki,
  Mod. Phys. Lett. {\bf A5} (1990) 1205;
  A. Hasenfratz, P. Hasenfratz, K. Jansen, J. Kuti and Y. Shen,
  Nucl. Phys. {\bf B365 } (1991) 79;
  E.A. Paschos and V.I. Zakharov,
  Phys. Lett. {\bf B272} (1991) 105; 
  J. Zinn-Justin,
  Nucl. Phys. {\bf B367} (1991) 105;
  P.M. Fishbane, R.E. Norton and T.N. Truong, 
  Phys. Rev. {\bf D46} (1992) 1768.
\bibitem{kn:BLL86} 
  W.A. Bardeen, C.N. Leung and S.T. Love, 
  Phys. Rev. Lett. {\bf 56} (1986) 1230;
  C.N. Leung, S.T. Love and W.A. Bardeen,
  Nucl. Phys. {\bf B273} (1986) 649.
\bibitem{kn:NSY89} 
  T. Nonoyama, T.B. Suzuki and K. Yamawaki, 
  Prog. Theor. Phys. {\bf 81} (1989) 1238.
\bibitem{kn:Higa84} 
  V.A. Miransky,
  Sov. J. Nucl. Phys. {\bf 38} (1983) 280; 
  K. Higashijima,
  Phys. Rev. {\bf D29} (1984) 1228.
\bibitem{kn:MNY89} 
  V.A. Miransky, T. Nonoyama and K. Yamawaki, 
  Mod. Phys. Lett. {\bf A4} (1989) 1409.
\bibitem{kn:Take89}
  T. Takeuchi,
  Phys. Rev. {\bf D40} (1989) 2697.
\bibitem{kn:BMSY87} 
  M. Bando, T. Morozumi, H. So and K. Yamawaki, 
  Phys. Rev. Lett. {\bf 59} (1987) 389.
\bibitem{kn:PR81}
  B. Pendleton and G.G. Ross, 
  Phys. Lett. {\bf 98B} (1981) 291.
\bibitem{kn:ABKMN90}
  K-I. Aoki, M. Bando, T. Kugo, M.G. Mitchard and H. Nakatani,
  Prog. Theor. Phys., {\bf 84} (1990) 683;
  K-I. Aoki, M. Bando, T. Kugo and M.G. Mitchard, 
  Prog. Theor. Phys.{\bf 85} (1991) 355.
\bibitem{kn:King90}
  S.F. King and S.H. Mannan,
  Phys.Lett. {\bf B241} (1990) 249; 
  F.A. Barrios and U. Mahanta,
  Phys. Rev. {\bf D43} (1991) 284.
\bibitem{kn:Suzu90b}
  M. Suzuki,
  Mod. Phys. Lett. {\bf A5} (1990) 1205;
  A. Hasenfratz, P. Hasenfratz, K. Jansen, J. Kuti and Y. Shen,
  Nucl. Phys. {\bf B365 } (1991) 79;
  E.A. Paschos and V.I. Zakharov,
  Phys. Lett. {\bf B272} (1991) 105; 
  J. Zinn-Justin,
  Nucl. Phys. {\bf B367} (1991) 105;
  P.M. Fishbane, R.E. Norton and T.N. Truong, 
  Phys. Rev. {\bf D46} (1992) 1768.
\bibitem{kn:Hill81}
 C.T. Hill, 
 Phys. Rev. {\bf D24} (1981) 691.
\bibitem{kn:STY90}
  S. Shuto, M. Tanabashi and K. Yamawaki,
  {\it in Proc. 1989 Workshop on Dynamical Symmetry Breaking, Nagoya, 
  1989,} eds. T. Muta and K. Yamawaki (Nagoya University, 1990);
  W.A. Bardeen and S.T. Love,
  Phys. Rev. {\bf D45} (1992) 4672;
  M. Carena and C.E.M. Wagner, 
  Phys. Lett. {\bf B285} (1992) 277.
\bibitem{kn:IKSY95}
  T. Itoh, Y. Kim, M. Sugiura and K. Yamawaki,
  Prog. Theor. Phys. {\bf 93} (1995) 417;
  K.-I. Kondo, 
  Nucl. Phys. {\bf B450} (1995) 251.
\end{thebibliography}
\end{document}